\pdfoutput=1
\documentclass[fleqn,10pt]{wlscirep}
\usepackage[utf8]{inputenc}
\usepackage[T1]{fontenc}
\usepackage[left]{lineno} % line numbers in the left margin
\usepackage{subcaption}
\usepackage{hyperref}

\usepackage{booktabs}
\usepackage{siunitx}
\usepackage{diagbox}
\title{A Data-Parsimonious Model for Long-Term Risk Assessments of West Nile Virus Spillover}

\author[1,*]{Saman Hosseini}
\author[2]{Lee W. Cohnstaedt}
\author[1]{Matin Marjani}
\author[1]{Caterina Scoglio}
\affil[1]{Department of Electrical and Computer Engineering,  
Kansas State University, Manhattan, KS, USA}
\affil[2]{Foreign Arthropod-Borne Animal Diseases Research Unit,  
USDA-ARS, Manhattan, KS, USA}

\affil[*]{Shosseini@ksu.edu}

\begin{abstract}
Many West Nile virus (WNV) forecasting frameworks incorporate entomological or avian surveillance data, which may be unavailable in some regions. 
We introduce a novel data-parsimonious probabilistic model to predict both the timing of outbreak onset and the seasonal severity of WNV spillover. 
Our approach combines a temperature-driven compartmental model of WNV with nonparametric kernel density estimation methods to construct a joint probability density function and a Poisson rate surface as function of  mosquito abundance and normalized cumulative temperature. 
Calibrated on human incidence records, the model produces reliable forecasts several months before the transmission season begins, supporting proactive mitigation efforts. 
We evaluated the framework across three counties in California (Orange, Los Angeles, and Riverside), two in Texas (Dallas and Harris), and one in Florida (Duval), representing completely different ecology and distinct climatic regimes, and observed strong agreement across multiple performance metrics.

\end{abstract}
\begin{document}

\flushbottom
\maketitle
% * <john.hammersley@gmail.com> 2015-02-09T12:07:31.197Z:
%
%  Click the title above to edit the author information and abstract
%
\thispagestyle{empty}

%\noindent Please note: Abbreviations should be introduced at the first mention in the main text – no abbreviations lists. Suggested structure of main text (not enforced) is provided below.

\section{Introduction}\
Arthropod-borne pathogens pose an increasing global health challenge, and WNV, a member of the family \textit{Flaviviridae} and genus \textit{Flavivirus}, was the most frequently reported domestically acquired arbovirus in the United States in 2023 \cite{kramer2008global,padda2025west}.
In the United States, WNV was first detected in New York in 1999 and quickly spread throughout the country in subsequent years and has since been responsible for thousands of confirmed human infections and hundreds of deaths \cite{kramer2019introduction,lanciotti1999origin,sejvar2003west}.
The transmission dynamics of WNV are complex, but involve primarily a cycle between mosquitoes of the \textit{Culicidae} family, particularly those of the \textit{Culex} genus \cite{turell2005update,troupin2016overview,weaver2010present}, and avian hosts, which serve as amplifier reservoirs. Occasionally, the virus spills over to mammals by mosquitoes, including humans and horses, which are considered dead end hosts due to their inability to maintain further transmission \cite{weaver2010present,hayes2005epidemiology}. 
Approximately 80\% of WNV infections are asymptomatic, while about 20\% manifest as West Nile fever.
Of these symptomatic cases, fewer than 1\% progress to neuroinvasive disease, such as meningitis, encephalitis, or acute flaccid paralysis.
Among people who develop neuroinvasive disease, the fatality rate in cases is approximately 10\% \cite{petersen2013west,gyure2009west}.

Despite the availability of equine vaccines \cite{ng2003equine,seino2007comparative}, no licensed vaccine is yet available for human use against WNV \cite{monath2001west,ulbert2019west}.
In the absence of human vaccination, population-level mitigation strategies, such as vector control programs and public health alerts, remain essential to limit WNV transmission \cite{cendejas2024vaccination,nasci2019reducing, bellini2014review}.
Accurate forecasting of WNV activity is integral to successful mitigation.
Such early warning insights allow vector control agencies to target larviciding and adulticiding resources where they will be most effective \cite{holcomb2021spatio,carney2008efficacy,mcmillan2021community}, while timely public health advisories boost community uptake of personal protective measures \cite{fonzo2024we}.
In this way, robust prediction transforms mitigation from reactive to proactive, improving both cost-effectiveness and health outcomes.

The early warning models developed for mitigation strategies employ different methodological frameworks and data inputs to forecast the risk of WNV spillover into humans \cite{barker2019models,keyel2021proposed}. 
In addition, these models can target nonhuman populations when predicting the risk of WNV and then translate those predictions into human risk. 
Models focusing on the mosquito population usually target prediction of mosquito infection rates or mosquito abundance, thus elucidating the future course of outbreaks in human populations. 
These models exploit the temporal delay between mosquito infection in mosquito populations and subsequent human outbreaks \cite{kilpatrick2013predicting,guptill2003early,selvey2014rainfall,kwan2010sentinel}. 
In this category, the models use a variety of non-surveillance data sources, together with mosquito surveillance data, depending on geographic location, environmental characteristics, and other contextual factors, to enhance predictive accuracy. 

Correlations between mosquito life-history traits—such as larval development rate, adult emergence, adult survival, and fecundity—and meteorological or hydrological variables have been well documented \cite{carrieri2014weather, valdez2017effects, shaman2002using}. In particular, the literature consistently shows that elevated temperatures occurring before or during the outbreak season increase the risk of transmission and spillover, regardless of geographic location \cite{wimberly2022integrated}. In contrast, the relationship between precipitation and WNV transmission is more variable and differs between regions \cite{hahn2015meteorological, wimberly2014regional, wimberly2022integrated}. Consequently, meteorological and hydrological data are often combined with mosquito surveillance data to model and predict transmission risk.
As an illustrative example, fine‐scale spatial and temporal models integrating mosquito abundance with temperature and precipitation data were applied to forecast the risk of WNV in northeast Illinois, where elevated air temperature and antecedent rainfall patterns explained approximately 80 \% weekly and 79 \% spatial variation in mosquito infection rates \cite{ruiz2010local}.
%A good example is Ruiz et al. (2010), who used fine-scale spatial and temporal models with mosquito, temperature, and precipitation data to forecast the risk of WNV in northeast Illinois. 
%Their model showed that the temperature and rainfall patterns explained most of the variation in infection rates \cite{ruiz2010local}. 
As a complementary example from a different region and data set, soil moisture, mosquito surveillance data, and 13 km grid weather variables were combined to model WNV infection in Suffolk County, NY, where the warm and dry April conditions best predicted the infection rate \cite{little2016development}. 
Furthermore, mosquito trap data of Emilia-Romagna (Italy) were used to fit to temperature-driven Bayesian mathematical models of \textit{Culex pipiens} and avian infection, estimating weekly human risk and linking it to reported cases with Poisson probability. 
The results show that 2018 had markedly higher mosquito/avian prevalence and approximately $8\times$ higher human transmission risk than previous seasons \cite{marini2020quantitative}.

In contrast, studies in arid and semi-arid regions have identified alternative key predictors, especially for hydrological parameters \cite{kovach2024irrigation,ukawuba2018association,shaman2010hydrologic}. 
In Lubbock County, TX, a nine-year CO$_{2}$-baited trap logistic regression analysis showed that wind speed was negatively associated with WNV-positive mosquito pools, whereas higher visibility, humidity, and dew point were each positively correlated with infection probability \cite{peper2018predictive}. 
Surprisingly, the model did not select temperature or precipitation as significant predictors, although both variables are commonly reported as important drivers of mosquito biology and the epidemiology of WNV.
Confirming that there are different hydrological predictive factors in the arid and semi-arid regions, a negative binomial model for Coachella Valley, CA, achieved its highest skill when potential evapotranspiration, rather than precipitation, was used as the main hydrological driver; In this arid setting, cooler, drier winters followed by warmer, wetter springs and cooler summers corresponded to higher infection rates \cite{ward2023spatially}.

In addition, other non-meteorological and non-hydrological factors relevant to mosquito infection or abundance have been also incorporated into these models alongside meteorological, hydrological, and mosquito surveillance data. 
For example, the inclusion of vegetation indices with climate and surveillance data was found to improve the understanding of mosquito population abundance and WNV infection \cite{britch2008satellite}.
Expanding on this, in Harris County, TX, time series models with mosquito surveillance, vegetation indices, and weather data were used to show that environmental variability and lagged winter temperature enabled WNV risk forecasts up to two months in advance \cite{poh2019influence}.
It is important to note that most early warning and predictive models based on surveillance data are tailored to specific local conditions and that predictors may not have consistent effects in different settings.
While previous studies have linked vegetation indices to improved WNV modeling, in Nassau County, NY, a negative association was identified between high normalized difference vegetation index (NDVI) and WNV incidence.
In the Bayesian spatio-temporal model developed in that study, which incorporated mosquito surveillance, meteorological, land use, and vegetation data, the highest WNV risk was detected in sparsely vegetated suburban areas, highlighting the significance of local environmental context \cite{myer2019spatiotemporal}.
This finding was confirmed by other work ~\cite{rosa2014early}.

Instead of focusing exclusively on mosquito populations, some models target other non-human hosts, such as birds or horses, using the corresponding surveillance data with non surveillance data to predict risk. 
Similarly to mosquito-based models, these approaches take advantage of the time lag between infection in the surveillance population and subsequent infection in humans.
For example, infections detected in dead birds or domestic pigeons have been shown to precede human cases by a measurable interval, providing empirical evidence for this lag \cite{eidson2001dead, chaintoutis2014evaluation}.
As a notable example of this approach, in northeast Florida advanced spatio-temporal models have been employed by integrating sentinel chicken seroconversion data with earth observation and climate variables to forecast transmission risk \cite{campbell2022spatiotemporal}.
Additionally, as another example, Manore et~al. developed a county-level early warning model that used county population size, publicly available bird population samples, and a prior-year human case indicator to predict risk in the following summer \cite{manore2014towards}.

%A notable example of this approach was provided by Campbell et al. (2022), in which advanced spatio-temporal models were developed using sentinel chicken seroconversion data, earth observation, and climate variables to predict transmission risk in northeast Florida \cite{campbell2022spatiotemporal}.
It should be noted that some models have used a combination of surveillance data, such as mosquito and sentinel chicken data, to establish the early warning framework \cite{chaskopoulou2013detection}.

There are models that predict the risk of WNV directly in the human population, using surveillance data such as mosquitoes, birds, or horses (or a combination of them) to train the model.
In a novel application, the accuracy of the forecast was improved using an ensemble-adjusted Kalman filter (EAKF), which allowed timely public health responses. 
A compartmental model was constructed and mosquito and human surveillance data was assimilated into it via the EAKF to directly predict the trajectory of the outbreak in the human population.
In addition to improved predictive performance, this model detected a seasonal pattern in mosquito bite rate \cite{defelice2017ensemble}.
The model was later enhanced by incorporating temperature \cite{defelice2018use}, and a subsequent study identified the impact of reporting delays in human and mosquito surveillance data on forecast accuracy \cite{defelice2019modeling}.

In a complementary line of work that moved from assimilation of surveillance data to combining surveillance data with climatic risk drivers, there are a variety of statistical models that frequently used this combination of data.
As an example in Illinois, a regression model showed that the most effective predictors of the model were fall precipitation, temperature of the previous week, and interactions between temperature and precipitation in different regions together with mosquito surveillance data \cite{karki2018assessing}.
Similarly, there are models that combine surveillance data with non-surveillance covariates to forecast the risk of WNV in the human population \cite{uelmen2020effects,davis2018improving}.

Although many models show strong performance in predicting outbreak timing, trajectory, and severity, their reliance on non-human surveillance data restricts applicability to locations with robust monitoring.
Moreover, as noted earlier, parameters that perform well at one site may fail at others, limiting the transferability between settings. 
A further practical constraint is prediction lead time: some models perform adequately over brief horizons but deteriorate at longer lead times, reducing their usefulness for early warning systems \cite{yi2024real}.
In contrast, several systems have produced skillful pre-peak forecasts and even projected seasonal totals weeks before the first reported human case, serving as de facto onset alerts, yet they still do not predict a formal onset interval \cite{defelice2017ensemble,defelice2018use,holcomb2023evaluation}.
Consequently, explicit prediction of a human spillover onset window remains uncommon. 
In parallel, most WNV early warning systems forecast monthly or seasonal case counts or risk levels; explicit prediction of spillover risk at daily resolution is similarly rare.
Finally, unmodeled reporting delays in mosquito and human surveillance can further degrade real-time skill \cite{defelice2019modeling}.

In this study, motivated by these gaps, we introduce a data-parsimonious probabilistic framework that combines a temperature-driven compartmental model of WNV with kernel density estimation \cite{silverman2018density} to predict human spillover risk.
Our contribution is an early-season predictive framework that is independent of non-human surveillance data, using only temperature and explicitly accounting for reporting delays, to forecast at daily resolution: (i) a calibrated temporal window for spillover onset and (ii) the subsequent day-by-day severity of spillover.

The framework is designed for early-season deployment: at the start of the year (or transmission season) we initialize the mosquito–temperature module with observed temperatures to date and a five-year climatology thereafter, yielding day-by-day guidance through season’s end.
Incorporating up-to-date temperature observations enables the model to capture recent climate and weather anomalies that influence mosquito propagation at the start of the season.
Beyond forecasting, it quantifies the threshold in mosquito abundance that governs spillover dynamics and reveals a persistent 30-year increase in high-risk days associated with gradual warming consistence with other works that show the effect of climate change on WNV \cite{fay2022west,keyel2021west}.
These features make the framework directly useful for early-warning operations and public-health planning.

\section{Material and methods}
This section outlines the study area, the disease transmission model, the data sources, the predictive modeling framework, the method used to capture long-term climatic trends in WNV risk and the validation procedures.
We begin by describing the key characteristics of the study region, followed by a presentation of the compartmental model developed to represent the dynamics of disease transmission. 
The subsequent subsection details the data and their sources.
We then introduce the predictive models used to forecast the timing of spillover onset and the severity of the risk.
After that we explain how the framework was applied to capture climatic trends in WNV risk, and conclude with a description of the validation approaches.
\subsection{Study location}
We evaluated the model in six counties in three states that represent distinct ecological settings. Orange, Riverside, and Los Angeles counties in southern California; Dallas and Harris counties in Texas; and Duval County in Florida.

From 2003 to 2024 the state of California recorded 8247 symptomatic human cases, 5082 neuroinvasive (62 \%) and 3113 non-neuroinvasive, along with 52 cases of other or unknown types and 406 fatalities.
Routine blood donor screening identified 860 additional asymptomatic infections during the same period \cite{CDPH2024WNV}.

WNV was first identified in Texas in 2002 and has since become established, with approximately 2,200 human cases reported through 2011.
In 2012, the state experienced its largest outbreak to date, with close to 1,900 infections.
That year, the Texas Department of State Health Services (DSHS) confirmed 1,868 cases, including 844 (45\%) cases of West Nile neuroinvasive disease (WNND) and 89 deaths, resulting in a case-fatality rate of 5\% \cite{murray2013west}. 

In Florida, WNV remains a persistent threat to public and animal health, where the state’s large equine industry places many breeding horses at risk of infection.
In Florida, a total of 460 WNV infections have been reported in people and 757 WNV infections in horses and other equines between 2001 and 2021.
Human and equine cases typically peak in the late summer and decline after November, aligning with increased mosquito activity \cite{cdc2022westnile,fdoh2022wnv}.

Additional information on the counties is provided in detail in {Appendix~A}.

\subsection{Disease model}
The compartmental model used for the WNV cycle mechanism is a $S_{H}E_{H}I_{H}R_{H} - Egg_{M}A_{M}S_{M}E_{M}I_{M} - S_{B}E_{B}I_{B}R_{B}$ model, where the subscripts $H$, $M$, and $B$ denote human, mosquito, and bird populations, respectively.
The compartments $S$, $E$, $I$, and $R$ represent susceptible, exposed, infectious (infected for humans) and recovered individuals.
In addition, the compartments $Egg$ and $A$ specifically denote the stages of the development of the oviposition and the aquatic stages of mosquitoes.
We have established our models using a single mosquito species (\textit{Culex tarsalis}) and a bird species of American crows (\textit{Corvus brachyrhynchos}).
In the model we have created, there are three categories of parameters related to birds, mosquitoes, and humans. 
The parameters for the human and bird population were obtained from the literature \cite{WHO2025WNV,brault2004differential,mclean2006west}.

In our model, the parameters of the mosquito population are all functions of temperature. 
These values were acquired from laboratory results as mechanistic coefficients \cite{shocket2020transmission}.
The model includes a parameter for the carrying capacity of the aquatic phase of mosquitoes ($K$) that was estimated by assimilating human case surveillance data using the Kalman filter.
Complete information and details about the compartmental model are reported in Appendix B. 

\subsection{Data}
Weekly reported human cases of WNV were obtained from the California Department of Public Health (CDPH), the Los Angeles County Department of Public Health, the Texas Department of State Health Services (DSHS), and the Florida Department of Health. Data were available and collected for the period 2006–2024 for Orange, Los Angeles, and Riverside counties, 2014–2024 for Dallas County, 2011-2024 for Harris County, and 2010-2024 for Duval County \cite{CDPH2024WNV, lacdph_home, texas_dshs_wnv,fdoh2022wnv}.
In addition, we used the average of the daily temperature data from 2006 to 2024 for the counties included in our predictions.
Temperature data were obtained from NASA’s POWER Project (Prediction Of Worldwide Energy Resource) \cite{nasapower}.

\subsection{Predictive models of spillover risk}
The risk of WNV spillover to humans was evaluated in two components: (i) predict the timing of the onset of the spillover and
(ii) predict the severity of the spillover events after the onset.
Mosquito abundance and temperature are strongly associated with WNV transmission dynamics \cite{shocket2020transmission,de2024effect,ciota2019role}; accordingly, our predictive framework centers on these two drivers.
Although additional ecological and environmental factors can contribute, the temperature was used as a proxy for broader environmental conditions.
In particular, we employ a normalized cumulative temperature metric to capture integrated thermal exposure, which better reflects the temperature dependence of mosquito development and WNV amplification than instantaneous temperature alone.
Thus, the predicted mosquito abundance of the compartmental model ($M$) together with the normalized cumulative temperature ($T$) serve as the principal input to our spillover risk models.

In the sections that follow, we detail how the two models were developed to predict spillover onset and spillover-severity risk.

\subsubsection{Long-term prediction of spillover onset} \label{sec: Long-term prediction of spillover onset}
The first part of our risk assessment focuses on predicting the timing of spillover onset each season; that is, our objective is to predict the precise time interval in which spillover onset occurs each year.
To achieve this, we obtained a bivariate predictive PDF, $f(M,T)$, which gives the probability density of spillover onset as a function of the observed values $(M,T)$, where $M$ is obtained from the compartmental model.
The predictive onset PDF determines a risk region in the MoT plane (values of $M$ on the horizontal and $T$ on the vertical axis) in which the probability of spillover onset is higher.
PDF is obtained by associating samples of $M$ from the compartmental model and $T$ with the time of spillover onset and estimating the density using the kernel density estimation method \cite{silverman2018density}.
An important consideration is that there is generally a temporal delay between the time of infection and the time of case reporting \cite{defelice2019modeling}, and this delay should be taken into account when constructing the predictive PDF.
The procedure for constructing the predictive PDF for spillover onset is presented in detail in Appendix C.

After obtaining the predictive PDF, we defined the risk contour $A_{\alpha}$ as the smallest subset of the MoT plane that contains an $\alpha$ fraction of the total probability mass:
\begin{equation}
  \iint_{A_{\alpha}} f(M,T)\,\mathrm dM\,\mathrm dT \;=\; \alpha ,
  \label{eq:alpha_region}
\end{equation}
and we refer to $\alpha$ as the confidence level.

\begin{figure}[htbp]
    \centering
    % Row 1: 0.5
    \begin{subfigure}{0.48\textwidth}
        \includegraphics[width=\linewidth]{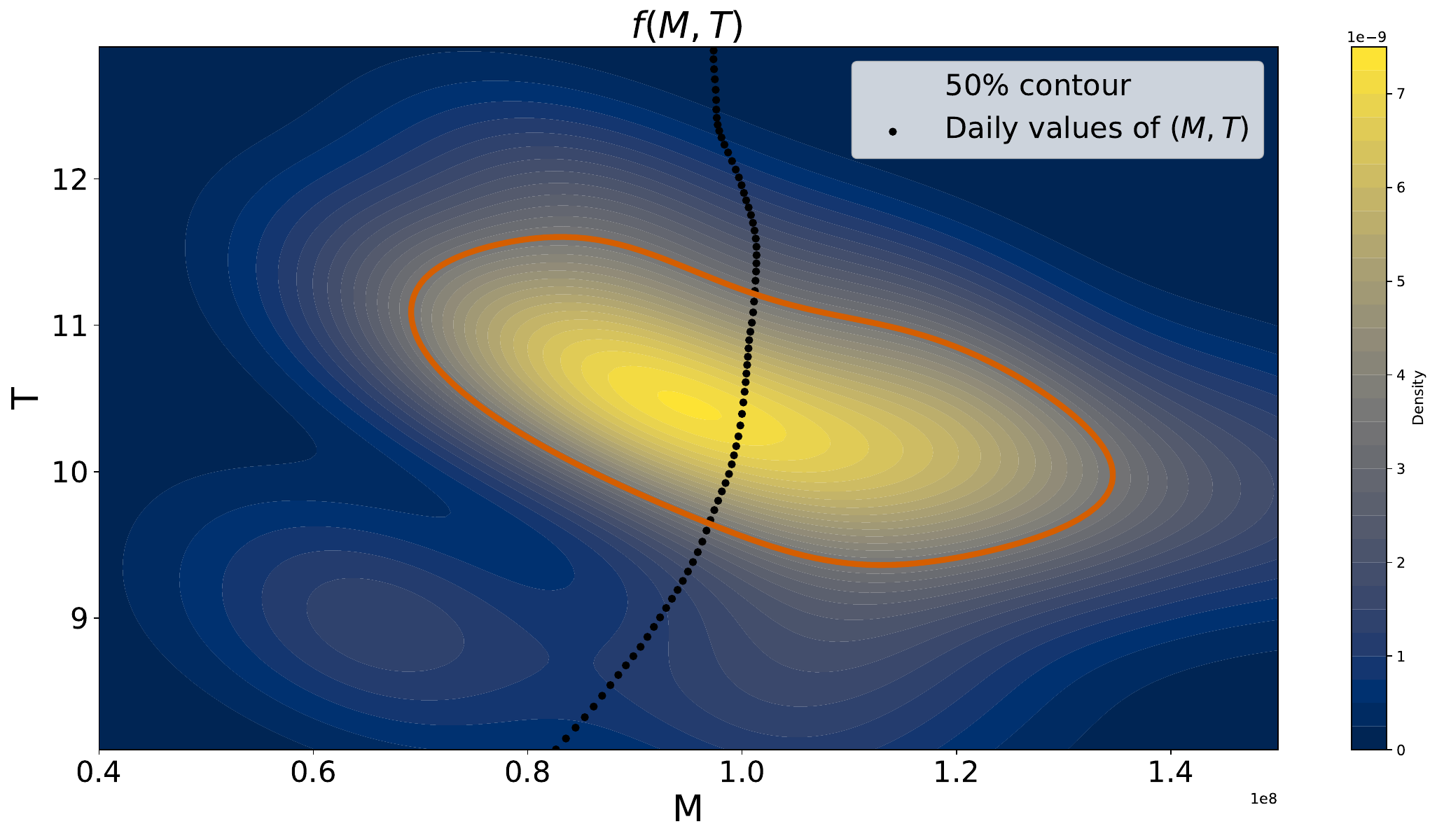}
        \caption{}
        \label{fig:pdf-contour-05}
    \end{subfigure}
    \hfill
    \begin{subfigure}{0.48\textwidth}
        \includegraphics[width=\linewidth]{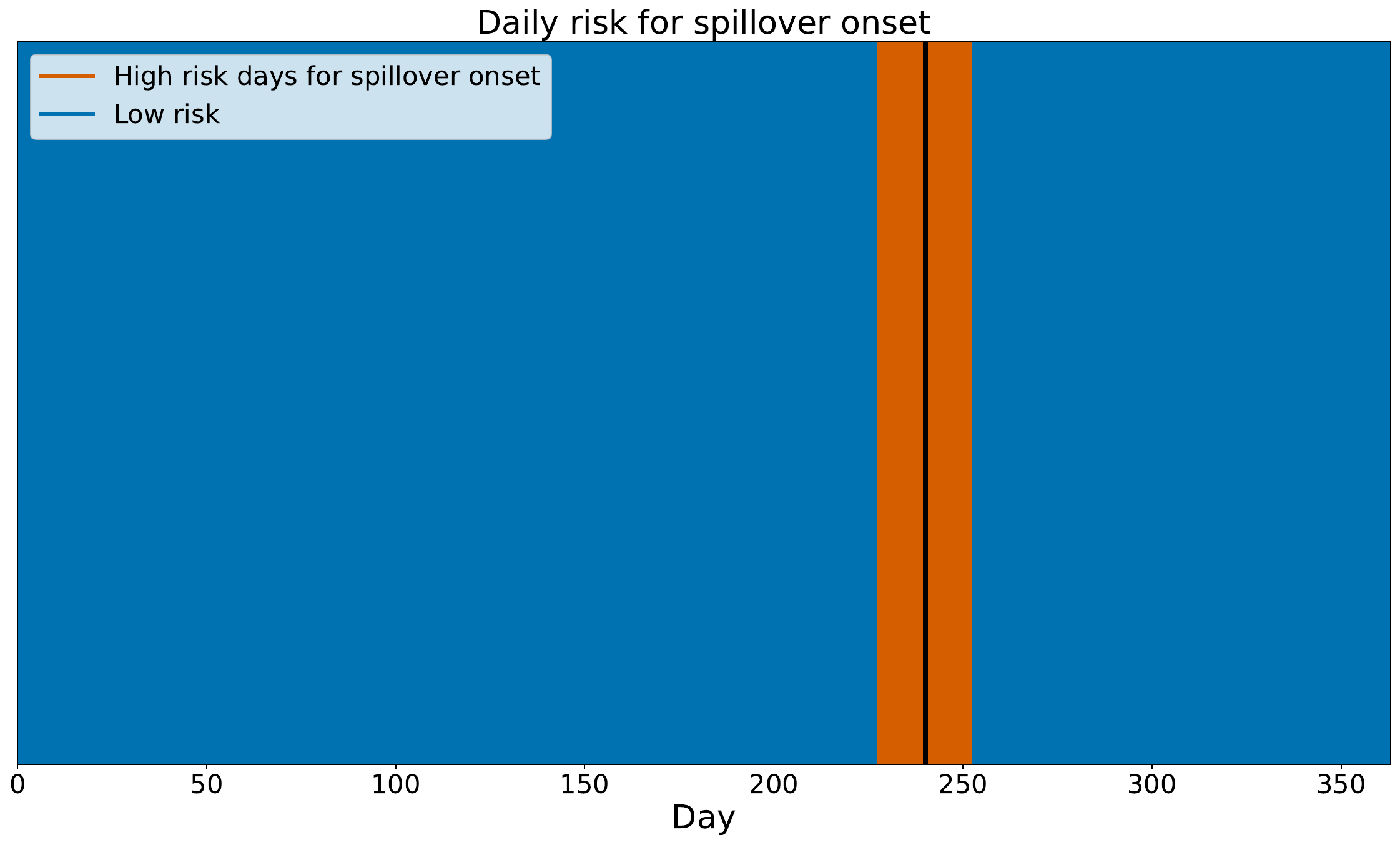}
        \caption{}
        \label{fig:timeline-05}
    \end{subfigure}

    % Row 2: 0.6
    \vspace{0.8em}
    \begin{subfigure}{0.48\textwidth}
        \includegraphics[width=\linewidth]{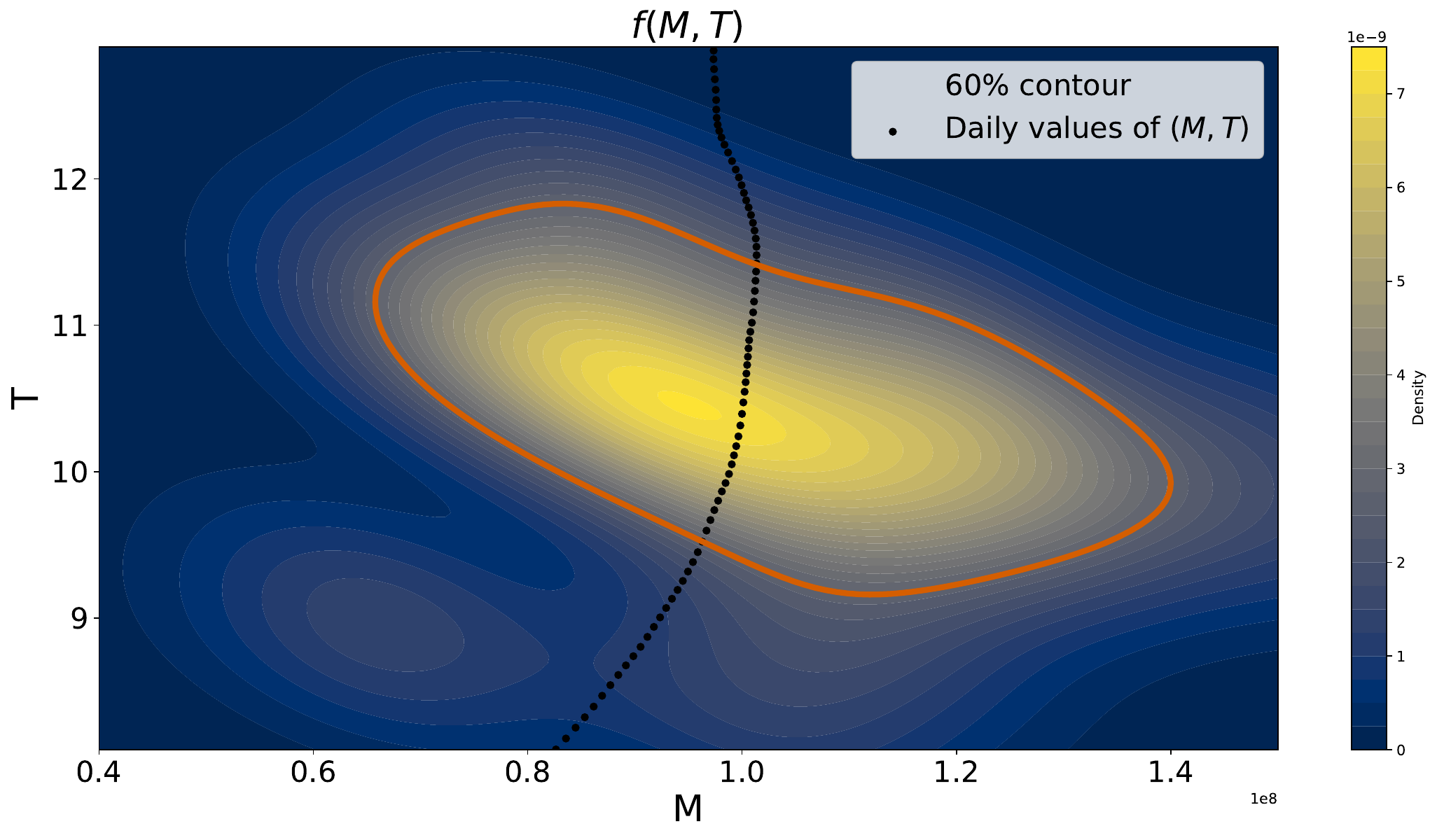}
        \caption{}
        \label{fig:pdf-contour-06}
    \end{subfigure}
    \hfill
    \begin{subfigure}{0.48\textwidth}
        \includegraphics[width=\linewidth]{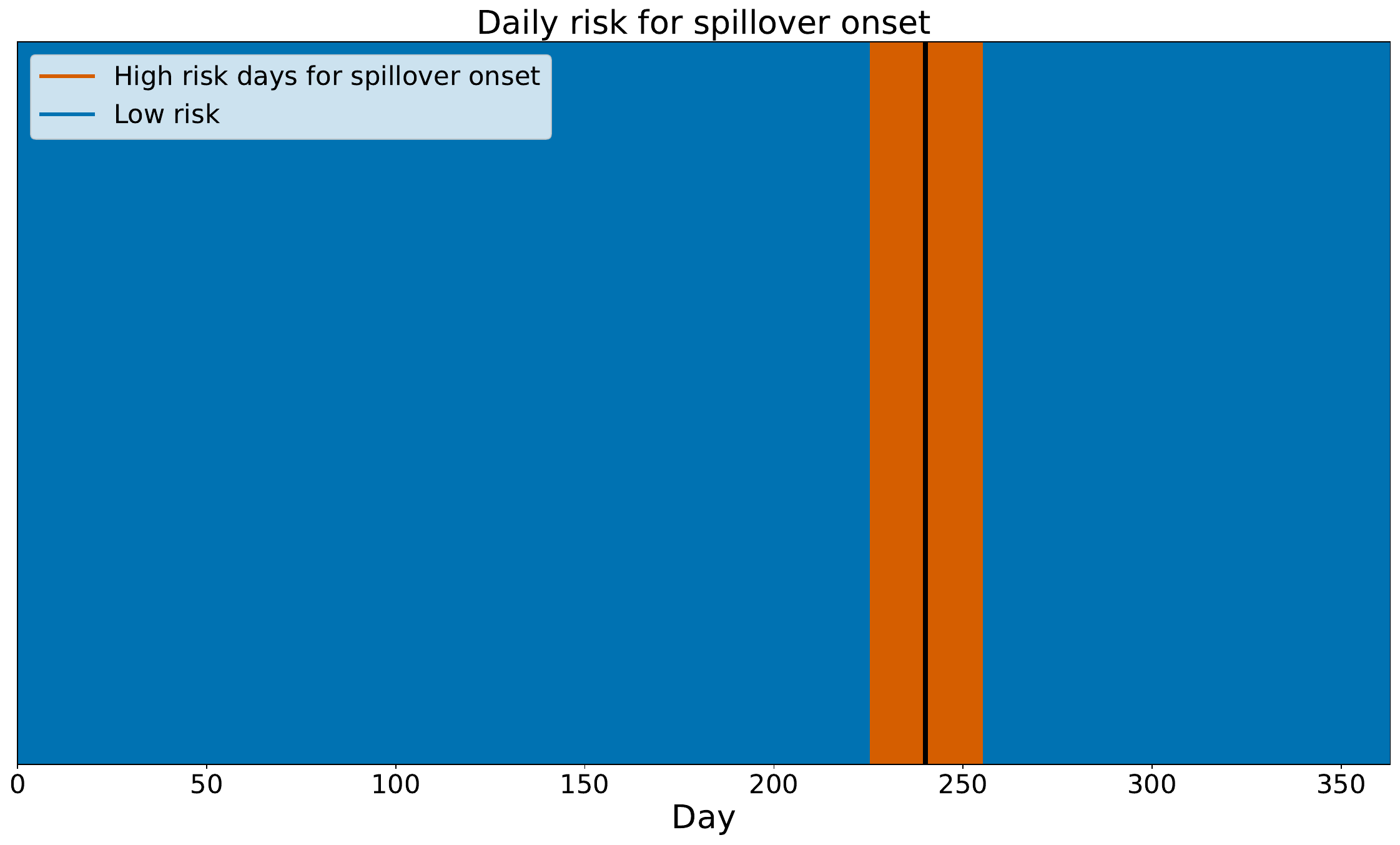}
        \caption{}
        \label{fig:timeline-06}
    \end{subfigure}

    % Row 3: 0.7
    \vspace{0.8em}
    \begin{subfigure}{0.48\textwidth}
        \includegraphics[width=\linewidth]{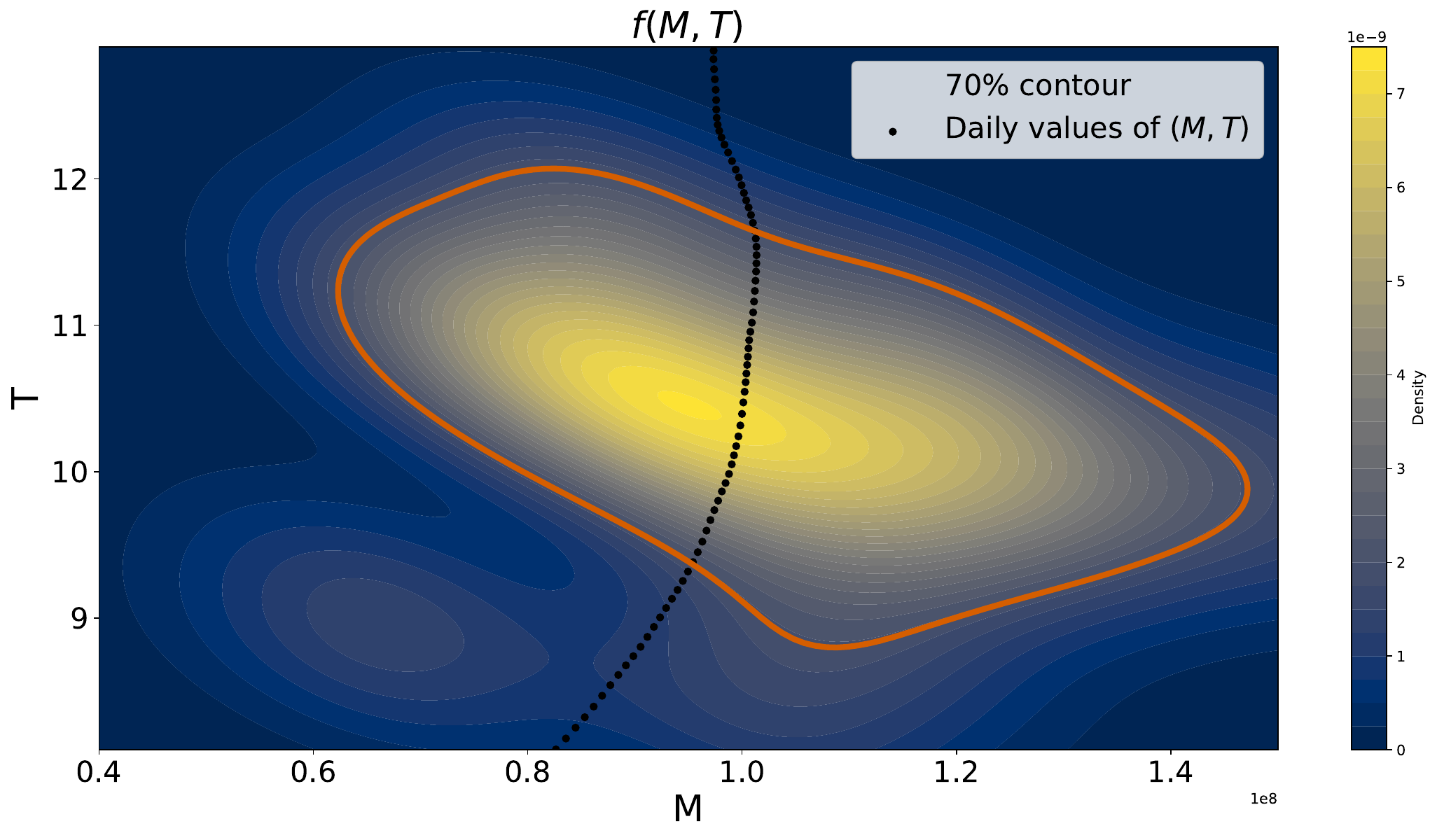}
        \caption{}
        \label{fig:pdf-contour-07}
    \end{subfigure}
    \hfill
    \begin{subfigure}{0.48\textwidth}
        \includegraphics[width=\linewidth]{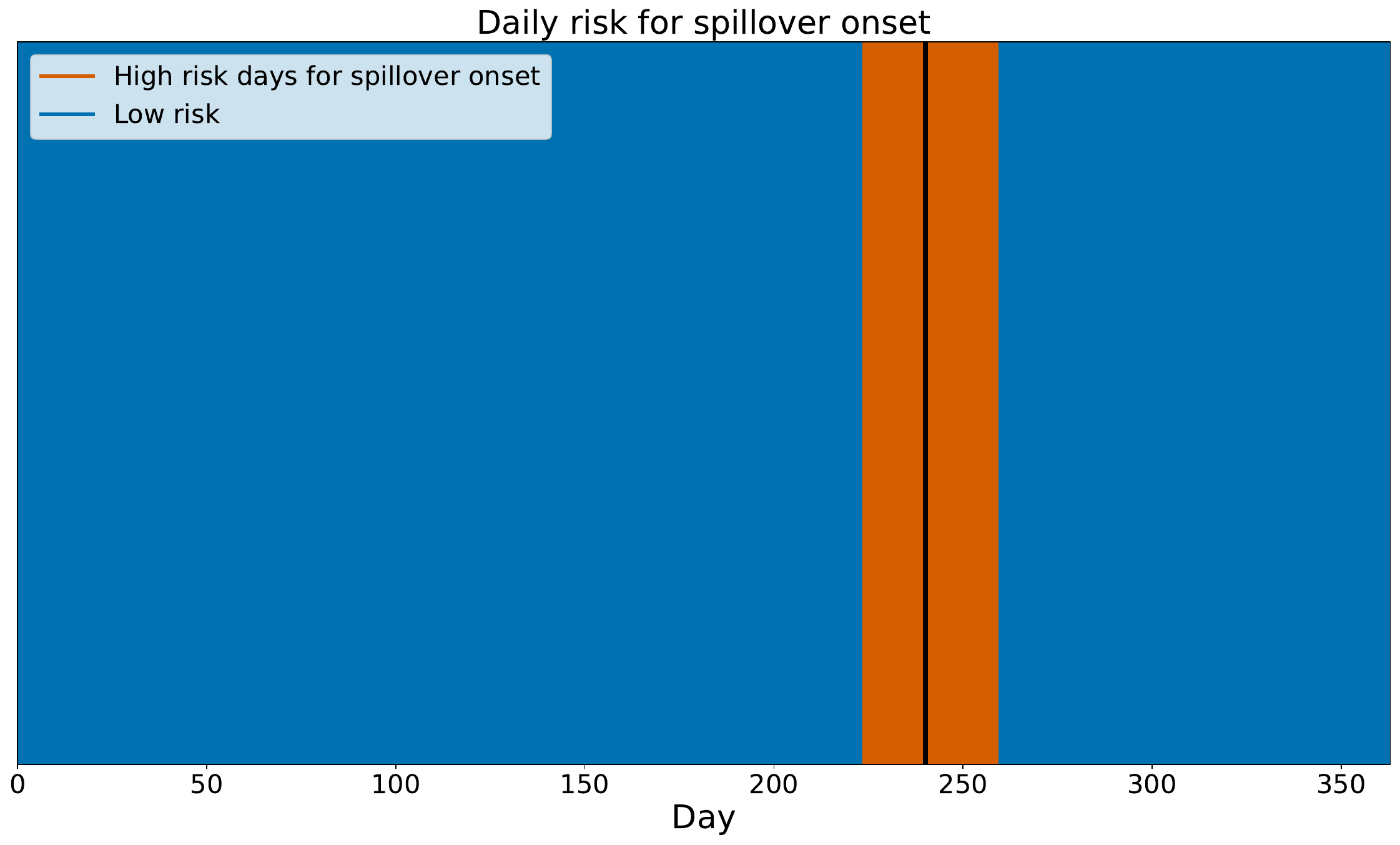}
        \caption{}
        \label{fig:timeline-07}
    \end{subfigure}

    \caption{
    Panels (a), (c), and (e) show the predictive PDF for spillover onset risk at different confidence levels ($\alpha = 0.5$, $0.6$, and $0.7$) for prediction in 2024, with the corresponding contour highlighted in orange.
    As illustrated in panels~(a), (c), and (e), the density of spillover events peaks in the lighter, high-probability regions of the joint \( (M,T) \) contour plots.
    This pattern indicates that the historical sample $(M,T)$ collected for 2006-2023 contains proportionally more observations with mosquito abundance \(M\) and normalized cumulative temperature \(T\) in those regions, conditions under which spillover was most likely to occur.
    Black dots represent daily $(M_i,T_i)$ values for 2024, and days where $(M_i,T_i)$ falls inside the risk contour of $\alpha$\% are considered “risky days.” Consequently, panels (b), (d), and (f) display the timeline of risky days across the year, as evaluated by the predictive PDF at each confidence level.
    Each pair of panels, (a,b), (c,d), and (e,f), illustrates a distinct confidence level \(\alpha\%\) together with the corresponding predicted interval for spillover onset.
    As anticipated, higher values of \(\alpha\) widen the prediction interval, whereas lower values narrow it.
    The PDF was derived from historical data (2006–2023), while the $(M,T)$ values for 2024 were predicted by running the compartmental model and average of temperature values.
    }
    \label{fig:pdf-timelines-paired}
\end{figure}

Using the predictive PDF and the corresponding risk contour (region) for a target year, we can assess the risk of spillover onset by evaluating the projected trajectory $\{(M_i,T_i)\}_{i=a}^{b}$ against the onset risk region, where $(M_i,T_i)$ corresponds to day $i$ in the interval from day $a$ to day $b$ (for example, $a$ can be the first day and $b$ the last day of the target year).
To predict the risk of spillover onset, $(M_i,T_i)$ values should be obtained using the predicted temperature values for the target year.
We also note that varying the confidence level $\alpha$ yields different contours, each delineating the values $(M,T)$ that correspond to a different probability of spillover onset risk.
Higher confidence levels yield broader contours and, consequently, longer time intervals.
Therefore, the choice of $\alpha$ should be guided by the appropriate metrics.
Figure~(\ref{fig:pdf-timelines-paired}) illustrates the effect of different confidence levels,their corresponding contours, on the length of the interval and the process of evaluating daily $(M,T)$ values with respect to these contours and the resulting time intervals.

\subsubsection{Eco-Epi Poisson model of spillover severity}\label{ratefunction}
After assessing the risk of spillover onset, we turned to assessing the potential severity of the resulting outbreak. 
The human spillover of WNV was modeled as an inhomogeneous Poisson process (Eco-Epi Poisson model) whose intensity is a function of abundance and normalized cumulative temperature ($M,T$).  
Let \(n(M,T)\) denote the number of human cases observed when the abundance of mosquitoes is \(M\) and the normalized cumulative temperature metric is \(T\).
We assume  
\begin{equation}
n(M,T) \;\sim\; \text{Poisson}\!\bigl(\lambda(M,T)\bigr),
\label{eq:poisson_model}
\end{equation}
where the rate surface \(\lambda(M,T)\) captures how expected spillover severity varies with environmental covariates \((M,T)\).
The surface of the spillover rate is estimated by assigning more weights to the region around \((M,T)\) in which the spillover has occurred.
To implement this, we apply kernel density estimation (KDE) methods to samples generated from the WNV compartmental model in times of spillover.
The procedure for constructing the Eco-Epi model rate surface \(\lambda(M,T)\), accounting for spillover reporting delay, has been explained in detail in Appendix~D.

Similarly to the spillover onset forecast, after obtaining the rate surface, we have to check the projected trajectory \(\{(M_i,T_i)\}_{i=a}^{b}\) against the rate surface of the Eco-Epi model to evaluate the severity of the risk for each day in the time interval from day $a$ to day $b$.
As indicated, for long-term prediction of risk (onset or severity of spillover), such as at the beginning of the year or before the outbreak season, it is necessary to predict the temperature conditions for the upcoming period to calculate the environmental variables \((M_i, T_i)\), which are then used to evaluate the rate surface and predict \(\lambda(M_i, T_i)\) in equation (\ref{eq:poisson_model}) or to check with the region of the spillover onset (\ref{eq:alpha_region}).
Following the prevailing practice in climate‐based disease modeling, we approximated future temperature by using the average of the last five years. 
Five-year smoothing windows are routinely employed to attenuate interannual variability: Lee et al. adopted the 5-year average daily temperature in the dengue risk projections for Jeju Island \cite{lee2018potential}; Sadeghieh et al. calculated monthly means using a moving 5-year average when simulating yellow fever transmission in Brazil \cite{sadeghieh2021yellow}; and the NASA - GISS global data set regularly disseminates the 5-year mean temperature change to emphasize the underlying trends \cite{hansen1987global}.
Collectively, these precedents support the five-year average as a pragmatic surrogate whenever high-resolution forecasts are unavailable. 
In practice, we combined real-time temperature observations for the target year prior to the start of the WNV season in the human population (i.e., day 140) with the five-year average temperature from previous years for the remaining days.

\begin{figure}
    \centering
    \includegraphics[width=0.9\linewidth]{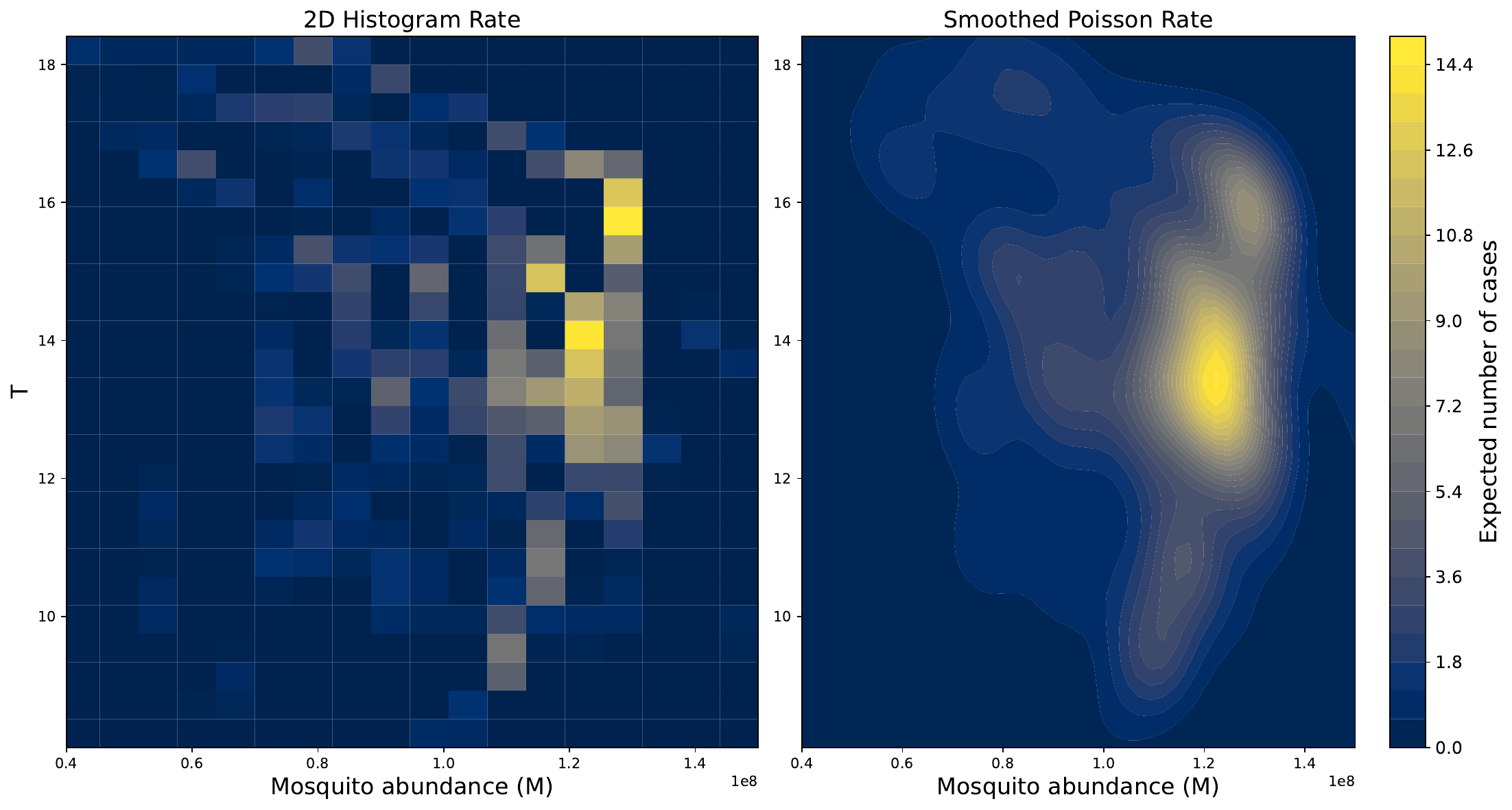}
    \caption{Left: Two-dimensional histogram of the estimated rate surface of the Eco-Epi Poisson model for Orange County, California, based on historical data from 2006 to 2023. 
    Each cell represents the estimated rate as a function of mosquito abundance \((M)\) and normalized cumulative temperature \((T)\).
    Right: Smoothed surface of the rate obtained by applying kernel density estimation to the collected data.
}
    \label{fig:Hist-rate}
\end{figure}

As a visual summary, Figure~\ref{fig:Hist-rate} shows a rate surface of the Eco-Epi Poisson model highlighting the zones of high, moderate, and negligible spillover risk ; brighter regions correspond to the environmental conditions most conducive to human infection.
Conceptually, the rate surface is a map over mosquito abundance \(M\) and normalized cumulative temperature \(T\) that identifies where the spillover risk is highest; its height, \(\lambda(M,T)\), is the expected number of infections (Poisson mean) given \((M,T)\).
It should be mentioned that we obtained both the PDF of the spillover onset and the rate surface of the Eco-Epi Poisson model with Scott-based kernel density estimation (KDE), a histogram-free method that avoids any binning artifacts (Appendix I).
However, for illustrative purposes, the 2-D histogram is shown in the left panel of Figure~\ref{fig:Hist-rate} to highlight the extra smoothing achieved by KDE relative to a simple histogram.

\begin{figure}
    \centering
    \includegraphics[width=0.9\linewidth]{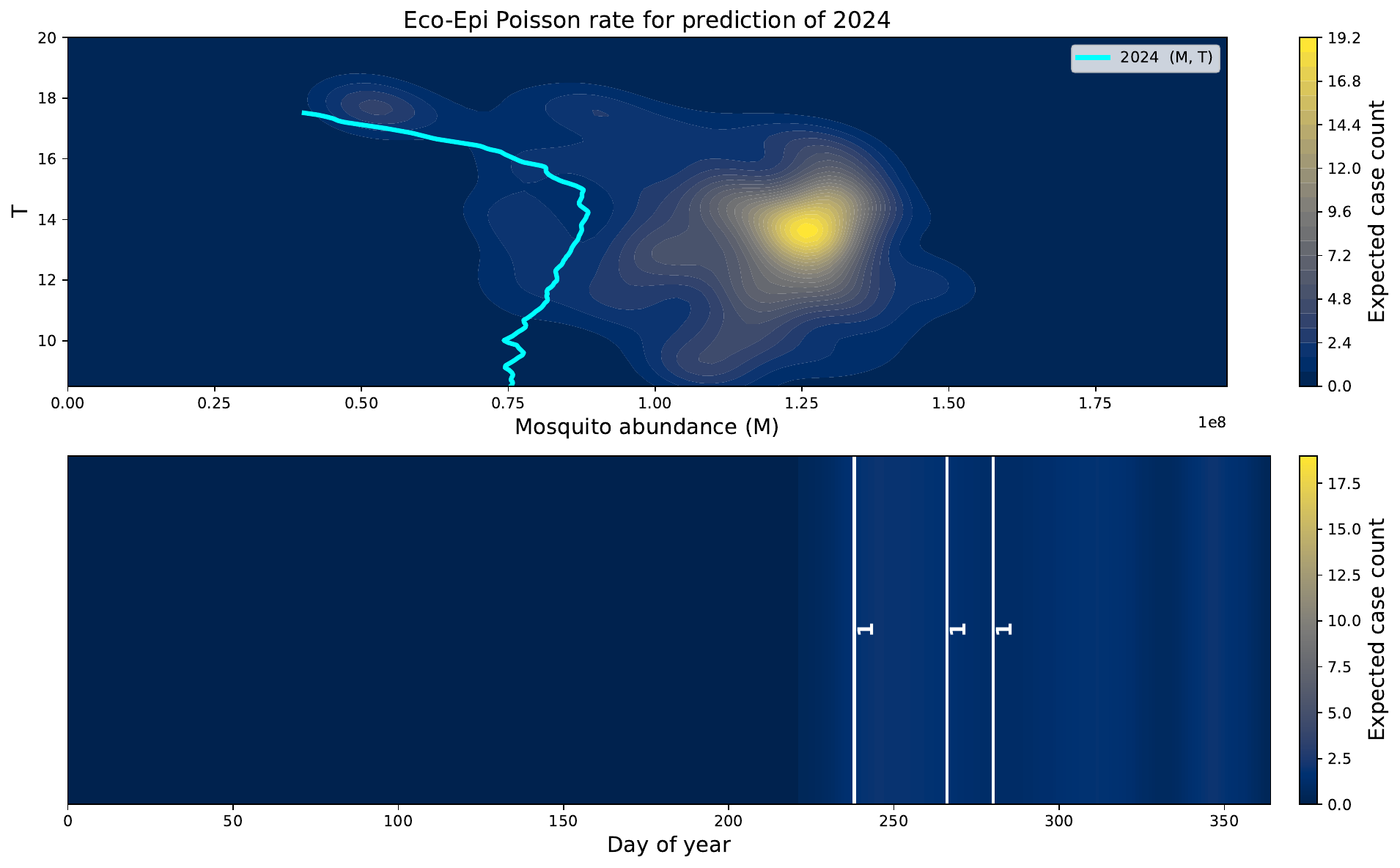}
    \caption{The top panel shows the smoothed rate surface of Eco-Epi Poisson model (obtained via kernel density estimation), along with the daily $(M,T)$ trajectory for 2024 (cyan curve).
    The bottom panel displays the predicted daily spillover risk for 2024 as evaluated from the rate surface after the observed spillover onset, along with the timing and count of actual reported human WNV cases in Riverside County, California.
    Reported case counts are displayed as bold white numerals positioned along the vertical lines indicating the reporting dates.
    The rate surface was estimated using historical data from 2006 to 2023.}
    \label{fig: Prediction_Severity_2024}
\end{figure}

As a practical example for the spillover severity forecast for 2024 in Riverside County, California, we obtained the rate surface of the Eco-Epi model.
Using the average temperature (from 2019-2023), we ran the compartmental model to calculate mosquito abundance \(M\), and we also estimated \(T\) for each day of the target year to obtain the corresponding $\{(M_i, T_i)\}_{i=0}^{364}$ values.
We then compared each pair of $(M_i, T_i)$ with the surface of the rate (the top panel of Figure~\ref{fig: Prediction_Severity_2024}).

Based on the result of this comparison, we predicted and plotted the daily risk using the corresponding value from the rate surface, which was visualized with an appropriate color. 
Figure~\ref{fig: Prediction_Severity_2024} illustrates the predicted daily rate values for 2024, shown as a temporal risk map.
The top panel shows the surface of the Eco-Epi Poisson rate as a function of the abundance of mosquitoes ($M$) and the normalized cumulative temperature ($T$).
The cyan curve represents the daily $(M_i,T_i)$ values for 2024, with its elevation above the surface indicating the predicted infection rate for each day.
The bottom panel translates this information into a temporal risk map, where each day of the year is shaded according to the corresponding Eco-Epi Poisson rate (i.e., the height of the trajectory curve) derived from the top panel.

\subsection{Long-term climatic trends and spillover risk} \label{Climate}

We investigated whether there is evidence supporting a relationship between long-term climatic trends and the risk of WNV spillover. 
To carry out this investigation, we first obtained the rate surface of Eco-Epi Poisson model using all years with human data (2006–2024), as explained in Section~\ref{ratefunction} and Appendix D. Consequently, we defined risky days as those in which the spillover risk \(\lambda(M,T)\) in Equation~\ref{eq:poisson_model} satisfies \(\lambda(M,T)\ge 1\).

Next, we examined long-term WNV risk by analyzing the model results over a 34-year period (1990–2024).
Although WNV was not reported in California before 2004, we nonetheless evaluated the potential risk during those earlier years in the absence of the pathogen. 
For each year from 1990 to 2024, we used the observed temperature to compute \(T\) and ran the compartmental model to compute \(M\). We then obtained \(\lambda(M,T)\) for each day and counted the number of risky days (days with  \(\lambda(M,T)\ge 1\)), resulting in the annual time series \(\{N_k\}\) for \(k = 1990,\ldots,2024\).
We analyzed this series to determine whether there was a statistically significant trend was present.

\subsection{Evaluation} \label{eval}
We evaluated the performance of the spillover onset prediction model from two perspectives: overall performance across all historical years and performance assessed separately for each individual year.
When predicting intervals for spillover onset, two factors are important: first, whether the predicted interval includes the actual time of spillover onset; and second, the length of the predicted interval. 
To account for both aspects, we define the metric \(\Tilde{A}_\alpha\) to evaluate the overall performance.
This metric captures the model’s ability to make correct predictions across multiple years, as well as the average length of the predicted intervals by the model at the specified confidence level \(\alpha\).
\begin{equation*}
    \Tilde{A}_{\alpha} =  \frac{\beta}{{\Tilde{L}_{\alpha}}},
\end{equation*}
in which $\beta$, the interval success rate, is the percentage of the year that includes the spillover onset time, and $\Tilde{L}$ is the average value of the predicted intervals for all historical years.

The second metric was defined as:
\begin{equation*}
    A_{\alpha} =\frac{ I_{\alpha}}{{L_{\alpha}}},
\end{equation*}
in which $I_{\alpha}$ indicates whether the prediction includes the spillover onset time and $L_{\alpha}$ counts for the length of the interval. 
We note that both $A_{\alpha}$ and $\Tilde{A}_{\alpha}$ receive values between zero and one as close to one as a more accurate prediction for the confidence level of $\alpha$. 

We must note that achieving a perfect value of \( A = 1 \) for both metrics is idealized and practically unattainable. 
To illustrate this, consider a scenario in which our method correctly predicts the interval such that it includes the actual spillover onset day throughout the year. 
If the predicted interval length is only 3 days, a relatively short window, the values of \( A_{\alpha} \) would be approximately 0.33. 
In other words, the value \( A_\alpha = 1 \) is only possible when the prediction interval exactly includes the true day of spillover onset, which corresponds to a point estimate.

\begin{figure}[htbp]
  \centering
  %-------------------------------------------------------------
  \begin{subfigure}[b]{0.45\linewidth}
    \includegraphics[width=\linewidth]{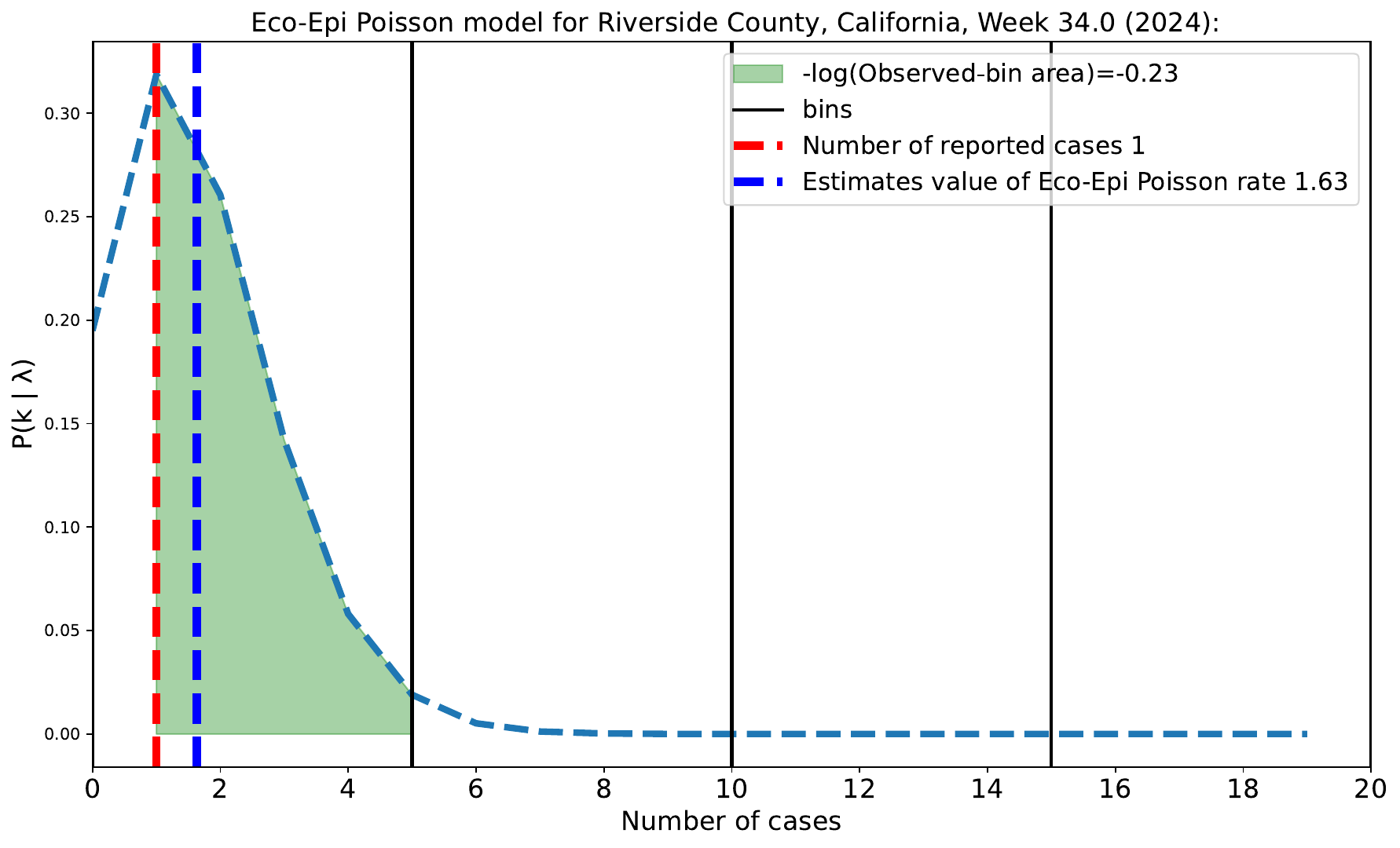}
    \caption{}
    \label{fig:logscore34}
  \end{subfigure}
  \hfill
  \begin{subfigure}[b]{0.45\linewidth}
    \includegraphics[width=\linewidth]{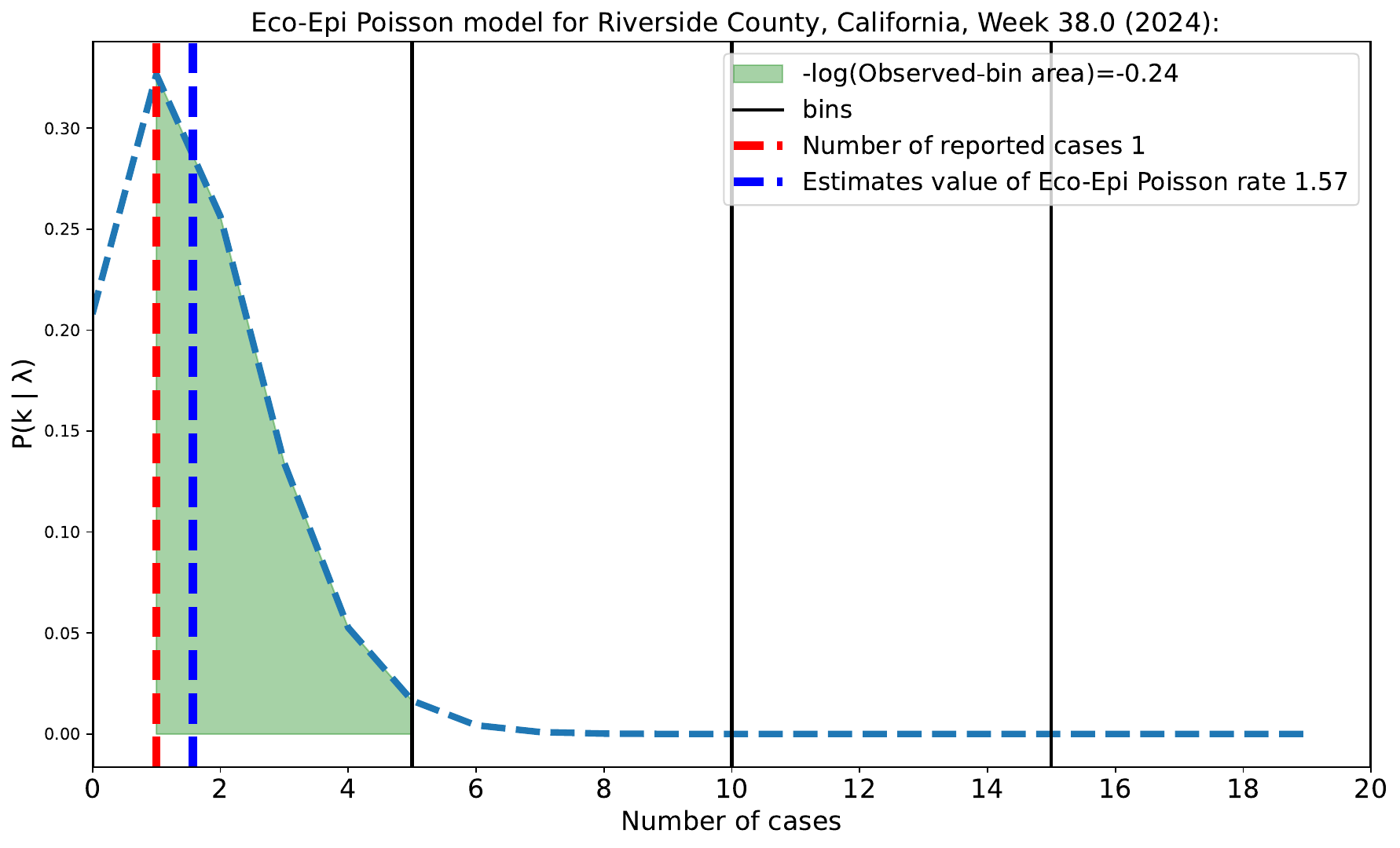}
    \caption{}
    \label{fig:logscore38}
  \end{subfigure}
  \hfill
  \begin{subfigure}[b]{0.45\linewidth}
    \includegraphics[width=\linewidth]{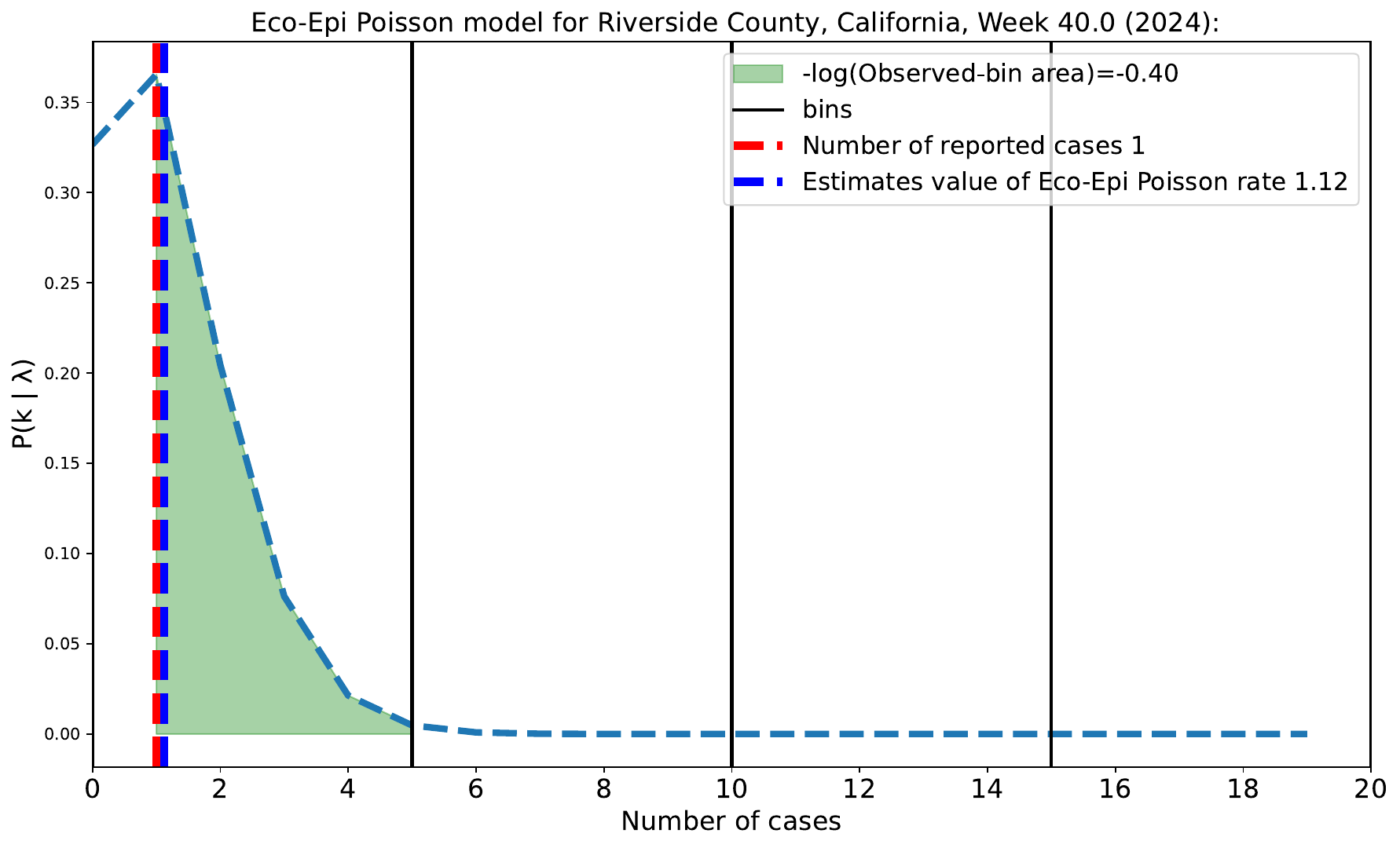}
    \caption{}
    \label{fig:logscore40}
  \end{subfigure}
  %-------------------------------------------------------------
  \caption{
  Logarithmic scores for the prediction of weeks with reported cases in Riverside County during 2024, based on leave-one-year-out cross-validation. Human cases were recorded in three weeks: 34 (day 238), 38 (day 266), and 40 (day 280). The Poisson distributions were generated using the predictive Eco-Epi rate surface calibrated on data from 2006 to 2023. For each of these weeks, the corresponding probability mass function is shown with epidemiological bins overlaid. The observed case counts are marked by a red vertical line, the predicted Eco-Epi rate are shown by blue vertical lines, and the bin containing the observation is shaded in green. The logarithmic score, computed from the probability mass within this bin, indicates high predictive accuracy, with values close to zero.
  It is noteworthy that the absolute differences between the predicted rates ($\lambda = 1.63$, $1.57$, and $1.12$) and the corresponding observed case counts are all less than one (i.e., $0.63$, $0.57$, and $0.12$, respectively).}\label{fig:logscore_riverside}
\end{figure}

Regarding the prediction of spillover severity, the predicted value of the rate surface of Eco-Epi model rate for each day has been considered as a point prediction for the number of people infected for a given day. 
Based on this assumption, to assess the performance of the model in prediction of risk severity, we considered three metrics. 
The first was the absolute value of the error (AVE), and the second is the normalized root mean square error (NRMSE). 
The third was the logarithmic score used by the Centers for Disease Control and Prevention (CDC) to evaluate the open prediction challenge to assess the ability to predict neuroinvasive disease of WNV, with further details provided in Appendix E.
We used the annual sum of logarithmic scores to assess the model's performance across different years.
Because \(\log P\) is non–positive, values closer to zero indicate better predictive performance \cite{holcomb2023evaluation}.
To clarify the scoring procedure, we noticed that for 2024 in Riverside County there were three weeks with reported cases (Figure \ref{fig: Prediction_Severity_2024}).
The rate surface predicted the rates of 1.63, 1.57 and 1.12 for these three weeks (weeks 34, 38, and 40), and consequently, 
Figure \ref{fig:logscore_riverside} shows the Eco-Epi Poisson model for these weeks.
The number of the reported cases have been shown by red dashed vertical lines on the Poisson pmf domain, and the black vertical lines are epidemiological bins. The prediction figures for all years and counties are provided in the supplementary materials (Supplementary 1.2.2 and 1.2.3).

We also used a null baseline defined as a Poisson process with a constant rate equal to the historical mean of weekly cases observed during the training period to test performance against chance. 
This formulation assumes no temporal dynamics or environmental influence and serves as a baseline to assess whether our proposed model provides meaningful predictive improvement beyond a constant rate assumption.
As a representative model from the literature and for direct performance comparison, we selected the negative binomial (NB) model. Trained on the number of new cases, the NB model achieved the highest score in the CDC prediction challenge evaluating the ability to predict neuroinvasive WNV disease \cite{holcomb2023evaluation}.

Also, for years without case reports, 2010 and 2022 in Riverside County and 2017, 2020, 2021, 2022, and 2023 in Duval County, we obtained the rate surface of the Eco-Epi model using other years with cases and predicted the risk for the mentioned years.
Then to evaluate these years, we calculated the seasonal sum of daily risk values and compared them with the corresponding sums from years with reported cases, in order to assess whether the total risk in no-case years was closer to the levels observed in years with few reported cases.

For the evaluation of long-term climatic trend, we applied three complementary statistical tests: linear regression, to detect and quantify a potential linear trend; Spearman rank correlation, to identify monotonic associations between time and value~\cite{sedgwick2014spearman}; and the Mann–Kendall test, which robustly detects monotonic trends without assuming linearity or normality~\cite{mcleod2005kendall}.
Together, these methods constitute a reliable framework for detecting and validating temporal trends.

\section{Results}
\subsection{Spillover onset}
To assess the predictive performance of the model in forecasting the timing of spillover onset, we adopted two strategies:
(1) Leave-One-Year-Out cross validation using data from 2006 to 2024, and (2) a retrospective forecasting approach, detailed below.
For each of them, we have used metrics explained in Section \ref{eval}.
Details of the leave-one-year-out cross-validation procedure and retrospective forecast for spillover onset have been explained in Appendix G.

\subsubsection{Leave-one-year-out cross validation for spillover onset prediction}
For Orange, Los Angeles, and Riverside counties, we performed a leave-one-year-out cross-validation covering 2006–2024. 
We repeated this cross-validation process for each year using various confidence levels ($\alpha = 0.50, 0.60, 0.70, 0.80, 0.90,0.95$ and $0.99$) (Eq. \ref{eq:alpha_region}).
In total, we generated 378 prediction intervals for spillover onset in various years and confidence levels (\(\alpha\)) for Orange, Riverside, and Los Angeles counties.
Among these predictions, 76\% (295 out of 385) included the actual spillover onset times within the predicted intervals.

In Orange County, for $\alpha = 0.50, 0.60, 0.70, 0.80, 0.90,0.95$ and $0.99$, the prediction interval included the real time of spillover onset in $73\%$, $73\%$, $73\%$, $74\%$, $84\%$, $89\%$, and $95\%$ of the years 2006 to 2024.
As explained in section (\ref{eval}) $\Tilde{A}_{\alpha}$ evaluate the overall performance of the predicted intervals.
The values of $\Tilde{A}_{\alpha}$ start at $0.028$ and end at $0.014$ for $\alpha = 0.50, 0.60, 0.70, 0.80, 0.90,0.95$ and $0.99$.
Table~\ref{tab:inverse-length-error-or}-a lists the lengths of the prediction intervals with their average for each level of confidence, the interval success rate $\beta$ and the overall performance $\Tilde{A}_{\alpha}$.
Table~\ref{tab:inverse-length-error-or}-b shows their associated accuracy ($A_\alpha$), an average of $A_\alpha$ for each level of confidence, and $\beta$ at various confidence levels for Orange County, California.

In Los Angeles, for $\alpha = 0.50, 0.60, 0.70, 0.80, 0.90, 0.95$ and $0.99$, the interval prediction captured the observation in $58\%$, $58\%$, $58\%$, $68\%$, $84\%$, $89\%$, and $95\%$ of the years 2006 to 2024.
The values of overall accuracy $\Tilde{A}_{\alpha}$ start from 0.0134 and end at 0.0072 (Supplementary 1.1.1).

For Riverside County, for $\alpha = 0.50, 0.60, 0.70, 0.80, 0.90, 0.95$ and $0.99$, the interval success rate $\beta$ is $59\%$, $65\%$, $76\%$, $76\%$, $76\%$, $88\%$, and $88\%$.
The values of overall accuracy $\Tilde{A}_{\alpha}$ start at 0.014 and end at 0.0077 (Supplementary 1.1.1).

%----------------------------------------------------------
%  Two compact subtables with a gap between them
%  Orange County: interval length vs. accuracy
%----------------------------------------------------------
\begin{table}[htbp]
  \centering
  \setlength{\tabcolsep}{4pt}

%------------------------------------------------------
%  (a) INTERVAL LENGTH  L_{α}  (days)  – UPDATED 0.999 COLUMN
%------------------------------------------------------
  \begin{minipage}{0.48\textwidth}
    \centering
    \footnotesize
    \caption*{(a) Interval length $L_{\alpha}$ (days)}
    \begin{tabular}{c|ccccccc}
      \toprule
      \diagbox{Year}{$\alpha$} & 0.50 & 0.60 & 0.70 & 0.80 & 0.90 & 0.95 & 0.99\\
      \midrule
      2006 & 29 & 35 & 39 & 44& 60 & 63 & 76\\
      2007 & 28 & 33 & 36 & 40 & 54 & 57 & 69\\
      2008 & 25 & 30 & 35 & 39 & 45 & 58 & 71\\
      2009 & 24 & 30 & 35 & 40 & 48 & 54 & 64\\
      2010 & 26 & 32 & 37 & 40 & 47 & 62 & 75\\
      2011 & 31 & 36 & 39 & 44 & 62 & 65 & 78\\
      2012 & 29 & 35 & 39 & 42 & 58 & 60 & 74\\
      2013 & 27 & 33 & 37 & 41 & 57 & 60 & 72\\
      2014 & 28 & 34 & 38 & 41 & 46 & 49 & 60\\
      2015 & 31 & 35 & 38 & 42 & 47 & 49 & 60\\
      2016 & 29 & 35 & 39 & 43 & 49 & 52 & 63\\
      2017 & 28 & 34 & 37 & 41 & 54 & 57 & 68\\
      2018 & 23 & 28 & 31 & 34 & 40 & 42 & 53\\
      2019 & 23 & 30 & 36 & 42 & 59 & 62 & 74\\
      2020 & 26 & 33 & 36 & 41 & 52 & 55 & 67\\
      2021 & 25 & 30 & 34 & 38 & 44 & 54 & 63\\
      2022 & 27 & 32 & 37 & 41 & 48 & 51 & 62\\
      2023 & 25 & 31 & 34 & 39 & 62 & 67 & 79\\
      2024 & 24 & 29 & 35 & 39 & 46 & 61 & 72\\
      \midrule
      \textbf{Avg($L_{\alpha}$)} & 26 & 32& 36& 40& 51 & 57 & 68\\
      \midrule
      \textbf{$\beta$(\%)} & 73 & 73 & 73& 74 & 84& 89 & 95\\
      \midrule
      \textbf{$\tilde{A}_{\alpha}$} & 0.028 & 0.022& 0.02 & 0.019& 0.016 & 0.015  & 0.014\\
      \bottomrule
    \end{tabular}
  \end{minipage}
  \hfill
%------------------------------------------------------
%  (b) ACCURACY  A_{α}  – UPDATED 0.999 COLUMN
%------------------------------------------------------
  \begin{minipage}{0.48\textwidth}
    \centering
    \footnotesize
    \caption*{(b) Accuracy $A_{\alpha}$}
    \begin{tabular}{c|ccccccc}
      \toprule
      \diagbox{Year}{$\alpha$} & 0.50 & 0.60 & 0.70 & 0.80 & 0.90 & 0.95 & 0.99\\
      \midrule
      2006 & 0.034  & 0.028 & 0.026 & 0.023 & 0.017 & 0.016 & 0.013\\
      2007 & 0.035  & 0.030 & 0.028 & 0.025 & 0.019 & 0.018 & 0.014\\
      2008 & 0      & 0      & 0      & 0    & 0    & 0.017 & 0.014\\
      2009 & 0.041  & 0.033 & 0.029 & 0.025 & 0.021 & 0.019 & 0.016\\
      2010 & 0.038  & 0.031 & 0.027 & 0.025 & 0.021 & 0.016 & 0.013\\
      2011 & 0.03   & 0.028 & 0.026 & 0.023 & 0.016 & 0.015 & 0.013\\
      2012 & 0.034  & 0.028 & 0.026 & 0.024 & 0.017 & 0.017 & 0.014\\
      2013 & 0.037  & 0.030 & 0.027 & 0.024 & 0.018 & 0.017 & 0.014\\
      2014 & 0      & 0    &   0    & 0    & 0.022  & 0.02  & 0.017\\
      2015 & 0.032 & 0.028 & 0.026 & 0.024 & 0.021  & 0.02 & 0.017\\
      2016 & 0.034 & 0.028 & 0.026 & 0.023 & 0.02   & 0.019 & 0.016\\
      2017 & 0.035 & 0.029 & 0.027 & 0.024 & 0.019  & 0.018 & 0.015\\
      2018 & 0     & 0      & 0      & 0     & 0    & 0     & 0.019\\
      2019 & 0     & 0      & 0      & 0     & 0.017& 0.016  & 0.014\\
      2020 & 0.038 & 0.0303 & 0.0278 & 0.024 & 0.019& 0.018 & 0.015\\
      2021 & 0     & 0      & 0      & 0      & 0   &   0    & 0\\
      2022 & 0.037 & 0.031 & 0.0263 & 0.024 & 0.021 & 0.02 & 0.016\\
      2023 & 0 .04 & 0.032 & 0.0278 & 0.026 & 0.016 & 0.015 & 0.013\\
      2024 & 0.041 & 0.034 & 0.0270 & 0.026 & 0.022 & 0.016 & 0.014\\
      \midrule
      \textbf{Avg($A_{\alpha}$)} & 0.027 & 0.022 & 0.02 & 0.018 & 0.016 & 0.015 &0.014\\
      \midrule
      \textbf{$\beta$(\%)} & 73 & 73 & 73& 74 & 84 &89  & 95\\
      \midrule
    \end{tabular}
  \end{minipage}

  \caption{Leave-one-year-out cross validation for spillover onset forecasts in Orange County, California: (a) interval prediction length and (b) accuracy \(A_{\alpha}\) as functions of confidence level \(\alpha\).}
  \label{tab:inverse-length-error-or}
\end{table}

\subsubsection{Retrospective forecast for spillover onset prediction}
We further assess the performance of the model prediction using retrospective predictions for the last two years.
The accuracy of the model was evaluated using the metric \( A_\alpha \) across confidence levels \(\alpha = 0.5, 0.55, 0.60, 0.65,\ldots, 0.95\) (Eq. \ref{eq:alpha_region}).
In Orange County, in 2023 the shortest interval containing the spillover onset time was at a confidence level of $0.5$ with a length of $26$ days and accuracy $A_\alpha=0.036$ and the longest interval obtained at a confidence level of $0.95$, its length $67$ days and $A_\alpha=0.015$.
For 2024 the shortest interval was obtained at $\alpha=0.5$ with $A_\alpha=0.036$ with a length of $28$ days, and the longest was obtained at the confidence level of $95\%$ with a length of $61$ days and $A_\alpha=0.016$ (Table \ref{tab:retro_or_2023_2024}).

Regarding Los Angeles, in 2023, the shortest interval was at a confidence level of 0.5 with a length of 41 days and accuracy $A_\alpha=0.0244$ and the longest interval obtained at a confidence level of 0.95, its length 62 days, and $A_\alpha=0.0161$.
For 2024 the shortest interval was obtained at $\alpha=0.5$ with $A_\alpha=0.0204$ with a length of 49 days and the longest was obtained at a confidence level of 95\% with a length of 66 days and $A_\alpha=0.0152$ (Supplementary 1.1.2).

For Riverside County, the shortest interval in 2023 was at a confidence level of 0.5 with a length of 36 days and accuracy $A_\alpha=0.0278$ and the longest interval obtained at a confidence level of 0.95, its length 68 days and $A_\alpha=0.0147$ .
For 2024 the shortest interval was obtained at $\alpha=0.5$ with $A_\alpha=0.0278$ with a length of 36 days and the longest was obtained at a confidence level of 95\% with a length of 68 days and $A_\alpha=0.0147$ (Supplementary 1.1.2).

\begin{table}[ht]
  \centering
  \setlength{\tabcolsep}{6pt}  % compact the columns a bit
  %--------------------------------------------------%
  %  LEFT SUB-TABLE : 2023
  %--------------------------------------------------%
  \begin{minipage}{0.48\textwidth}
    \centering
    \caption*{(a) Orange County — prediction year 2023}
    \begin{tabular}{ccc}
      \toprule
      $\alpha$ & Length (days) & $A_{\alpha}$ \\
      \midrule
      0.50 & 26 & 0.036 \\
      0.55 & 30 & 0.033 \\
      0.60 & 31 & 0.032 \\
      0.65 & 32 & 0.031 \\
      0.70 & 35 & 0.029 \\
      0.75 & 55 & 0.018 \\
      0.80 & 58 & 0.017 \\
      0.85 & 60 & 0.017 \\
      0.90 & 63 & 0.016 \\
      0.95 & 67 & 0.015 \\

      \bottomrule
    \end{tabular}
  \end{minipage}
  \hfill
  %--------------------------------------------------%
  %  RIGHT SUB-TABLE : 2024
  %--------------------------------------------------%
  \begin{minipage}{0.48\textwidth}
    \centering
    \caption*{(b) Orange County — prediction year 2024}
    \begin{tabular}{ccc}
      \toprule
      $\alpha$ & Length (days) & $A_{\alpha}$ \\
      \midrule
      0.50 & 28 & 0.036 \\
      0.55 & 32 & 0.031 \\
      0.60 & 32 & 0.031 \\
      0.65 & 34 & 0.029 \\
      0.70 & 51 & 0.020 \\
      0.75 & 53 & 0.019 \\
      0.80 & 54 & 0.019 \\
      0.85 & 56 & 0.018 \\
      0.90 & 58 & 0.017 \\
      0.95 & 61 & 0.016 \\

      \bottomrule
    \end{tabular}
  \end{minipage}

  \caption{Retrospective-forecast interval length and accuracy metric ($A_{\alpha}$) for Orange County, California, at several confidence levels.}
  \label{tab:retro_or_2023_2024}
\end{table}

\subsubsection{Predictive PDFs for Spillover onset}\label{Predictive PDFs}
Between the three counties, Orange County shows the smallest variation in the averaged predicted interval length (${Avg}(L_{0.99})-{Avg}(L_{0.5})$=68-26=42 days), compared to Riverside (113-42=71 days) and Los Angeles County (128-43=85 days). 
Predictive spillover PDFs support this result, indicating that the variability of the normalized cumulative temperature ($T$) (and consequently the temperature) in Orange County is lower than in Riverside and Los Angeles.
Consequently, the predictive PDF for spillover onset indicates a shorter temperature range in Orange County (min. 5.8 and max. 14.6).
In contrast, the onset occurs across a wider thermal window in Riverside (min. 3.5 and max. 16.2) and Los Angeles (min. 3.7 and max. 15.1). 
In addition, the minimum value of \(T\) in Orange County exceeds those of Riverside and Los Angeles, while its maximum \(T\) is lower than the corresponding maxima in the other two counties (Supplementary 2).

\subsection{Spillover severity}
\subsubsection{Rate surface of spillover}\label{sec: rate}
\begin{figure}
    \centering
    \includegraphics[width=0.7\linewidth]{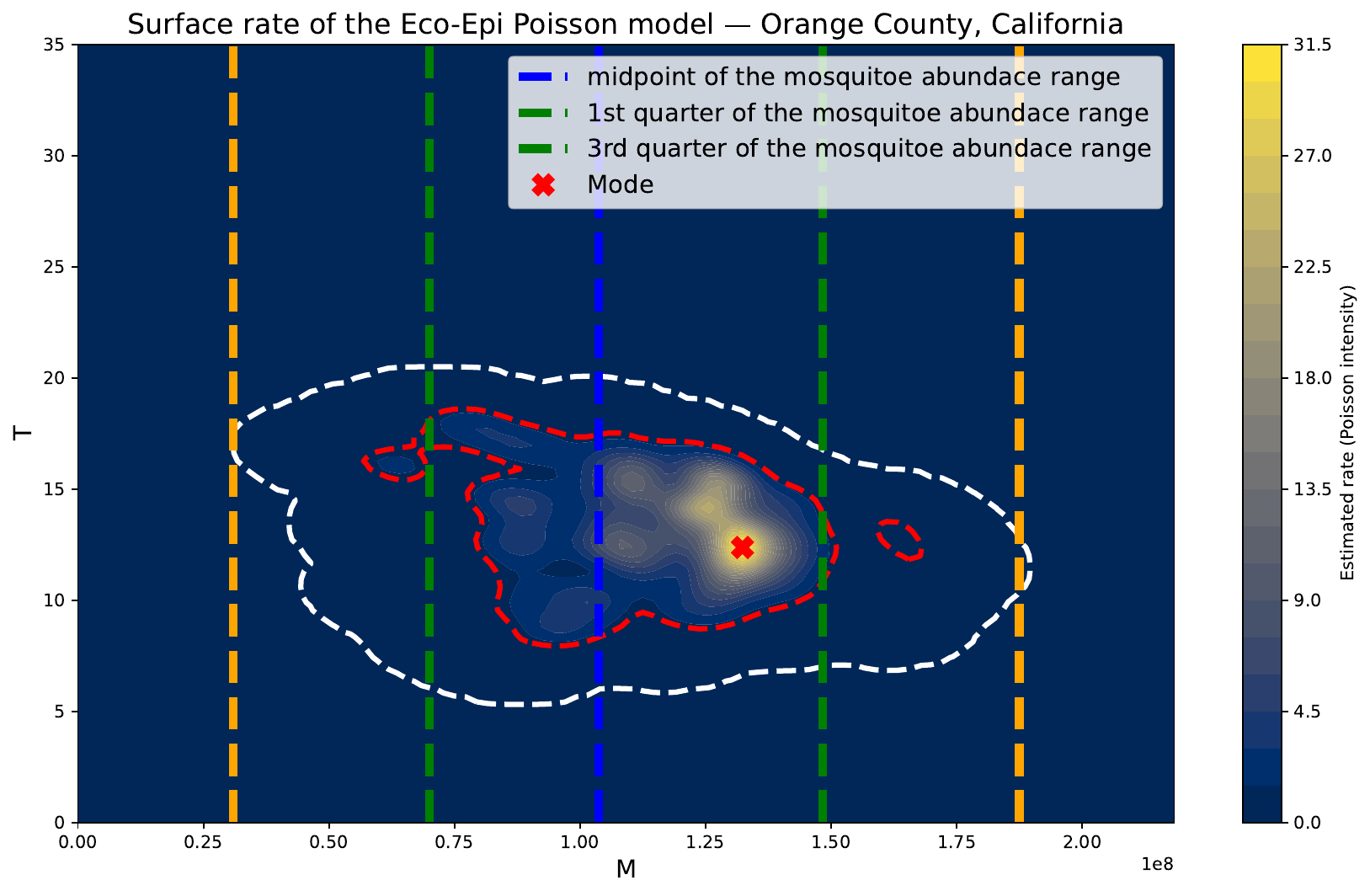}
    \caption{
Eco-Epi Poisson model rate surface estimated from historical human WNV cases using a KDE-based method for Orange County, California (2006–2024). The white dashed contour encloses the region with nonzero rate, while the red dashed contour highlights the area where the rate exceeds one. Vertical dashed lines mark the first quartile (green), midpoint (blue), third quartile (green), and maximum value (orange) of the mosquito abundance range ($M$). The red marker indicates the mode of the estimated Eco-Epi Poisson rate surface, showing that the highest spillover intensity occurs at intermediate mosquito densities rather than at extreme values.}

    \label{fig:intermediate_M_OR}
\end{figure}

If we consider the rate surface of the Eco-Epi Poisson model (Eq(\ref{eq:poisson_model})), for nonzero rate values (inside the region of the white dashed contour), this surface for Orange County shows three main characteristics (Figure \ref{fig:intermediate_M_OR}).
First, the spillover rate exceeds unity (red dashed contour) between the minimum mosquito abundance ($M$) and approximately its first quartile (denoted by the orange dashed line for the minimum value and the green dashed line for the first quartile in Figure~\ref{fig:intermediate_M_OR}), thereby defining a threshold in $M$ for the onset of the spillover.
Second, within an intermediate window of $T$, the spillover rate increases with $M$ until it reaches a maximum.
Third, beyond the peak, further increases in mosquito abundance $M$ do not yield commensurate gains in the predicted spillover rate; instead, the rate declines sharply.
A similar pattern is observed for other counties (Supplementary 1.2.1).\\

To evaluate the predictive performance of the Eco-Epi Poisson model in forecasting outbreak severity, we employed two strategies, consistent with those used for spillover onset: (1) leave-one-year-out cross-validation and (2) retrospective forecasting.
The procedures for cross-validation and retrospective forecast of spillover severity are described in detail in Appendix H.

\subsubsection{Leave-one-year-out cross validation for spillover severity prediction}
In this section, we applied the one-year-out cross-validation policy across all study years, and for each excluded year, we obtained the risk of spillover by predicting the spillover rate $\lambda$ of the Eco-Epi Poisson model (Eq.~\ref{eq:poisson_model}) for each day of the outbreak season in that year.
Predictions were generated for all years 2006-2024 in Orange, Riverside and Los Angeles counties in California; 2014–2024 in Dallas and 2011-2024 in Harris counties in Texas; and 2014–2024 in Duval County, Florida. 
Consequently, to check the performance of the model, we evaluated the results using the metrics explained in Section \ref{eval} and compared the results of the Eco-Epi model with those of the null baseline and negative binomial model.

For Riverside County, the annual sums of logarithmic scores for the Eco-Epi Poisson model showed a better performance compared to both the NB and the null models (Supplementary 1.2.2.13).
Likewise, during 2006–2024, the Eco-Epi Poisson model outperformed the null baseline in terms of NRMSE and generally surpassed the NB model, with two exceptions, 2009 and 2011 (0.66 vs 0.43 and 0.46 vs 0.54) (Supplementary 1.2.2.14).
Among the 149 weekly predictions, only 3 weeks (3 weeks in 2015) were identified as outliers, characterized by a logarithmic score of \(-10\).
For 138 of the 149 weeks (\(92\,\%\)), the observed values fell within the bin with the highest predicted density (meaning the best possible logarithmic score)(Supplementary  1.2.2.2).
Based on AVE values, \(92.62\,\%\) of all weekly predictions for Riverside County have an error of fewer than 4 cases, while \(4.7\,\%\) exhibit an error greater than or equal to 4.
The distribution of errors is as follows: error \(<1\): \(42.95\,\%\); error \(<2\): \(73.15\,\%\); error \(<3\): \(92.62\,\%\); error \(<4\): \(95.3\,\%\); and error \(\ge 4\): \(4.7\,\%\).

In Orange County, according to the logarithmic score, the Eco-Epi model outperformed the null baseline in all years except 2018 (-4.57 vs. -4.49).
Furthermore, Eco-Epi model outperformed the NB model in all years except in 2018, 2019, and 2021 (-4.57 vs -3.47, -2.36 vs -2.18 and -1.33 vs -1.12).
In terms of the NRMSE metric, the Eco-Epi model outperformed the null baseline in all years.
The Eco-Epi model performed comparably with the NB model in years with low case counts, but in years with high case counts it substantially outperformed both the NB model, for example, 2008 (1.6 vs 6.2 and 3.9), 2012 (0.6 vs 5.6 and 1.4), 2014 (1.7 vs 43.0 and 17.8) and 2015 (2.1 vs 10.5 and 3.6) (Supplementary 1.2.2.13 and 1.2.2.14).
Among the 155 weeks of perdition, only 6 weeks (\(0.04\,\%\)) were identified as outliers, characterized by a logarithmic score of \(-10\) (Supplementary 1.2.2.4).
In 133 of the 155 weeks (\(86\,\%\)), the observed values fell within the bin with the highest predicted density (Supplementary 1.2.2.4).
Based on AVE values, \(85.81\,\%\) of all weekly predictions for Orange County have an error of fewer than 5 cases, while \(14.19\,\%\) exhibit an error greater than or equal to 5.
The detailed distribution of the errors is as follows: error \(<1\): \(17.42\,\%\); error \(<2\): \(32.26\,\%\); error \(<3\): \(50.32\,\%\); error \(<4\): \(73.55\,\%\); error \(<5\): \(85.81\,\%\); and error \(\ge 5\): \(14.19\,\%\).

For Los Angeles County, the Eco-Epi Poisson model outperformed the null baseline in all years except 2009 ($-5.31$ vs. $-4.6$) based on the annual sum of the logarithmic score. 
In addition, the Eco-Epi model outperformed the NB model in 79\% of the years based on the logarithmic score and in years 2008, 2012, 2014, 2015,2017, 2022 (-58.16 vs. -39.79, -53.71 vs. -35.23, -75.30 vs. -43.99, -82.23 vs. -68.95, -103.70 vs -64.28, and -8.15 vs. -7.73).
In terms of NRMSE, the Eco-Epi model performed better than the null baseline in all years except 2006 and 2009 ($0.7$ vs. $0.4$ and $0.6$ vs.\ $0.5$) (Supplementary~1.2.2.13 and~1.2.2.14).
According to NRMSE, the Eco--Epi model outperformed NB in 15 of 19 years (79\%) between 2006 and 2024.
The years with lower performance were 2008, 2012, 2013, and 2017 ( 2.5 vs 1.5, 1.6 vs 0.9, 1.1 vs 0.9 and 3 vs 2.8).
A total of 254 weekly predictions were generated as part of the leave–one–year–out cross-validation, corresponding to weeks with reported cases from 2006 to 2024.
The predictions for 16 weeks (6.2\%) were identified as outliers, characterized by a logarithmic score of \(-10\) (Supplementary 1.2.2.6).
Based on AVE values, 70.08\% of all weekly predictions for Los Angeles County have an error of fewer than 5 cases, while 29.92\% exhibit an error greater than or equal to 5. The detailed distribution of the errors is as follows: error \(<1\): 17.72\%; error \(<2\): 33.07\%; error \(<3\): 45.28\% ; error \(<4\):61.81\%; error \(<5\): 70.08\%; and error \(\ge 5\): 29.92\%.

For Duval County, based on the annual sum of logarithmic scores, the Eco-Epi Poisson model outperformed the null baseline and the NB models in all years.
According to the NRMSE metric, the Eco-Epi model performed better in all years compared to the null baseline and the NB model.
Throughout all prediction weeks, there was no outlier prediction. 
In fact, 48 out of 49 weekly predictions ($98\%$) fell within the bin with the highest predicted density (Supplementary 1.2.2.11 and 1.2.2.12).
Based on AVE values, $97.96\%$ of all weekly predictions for Orange County had an error of fewer than 5 cases, while only $2.04\%$ exhibited an error greater than or equal to 5.
The detailed distribution of the errors is as follows: error $<1$: $61.22\%$; error $<2$: $91.84\%$; error $<3$: $95.92\%$; error $<4$: $95.92\%$; error $<5$: $97.96\%$; and error $\geq 5$: $2.04\%$ (Supplementary 1.2.2.13 and 1.2.2.14).

For Dallas County, the Eco-Epi model outperformed both the null baseline and the negative binomial model in terms of logarithmic score (Supplementary 1.2.2.13 and 1.2.2.14). 
Based on the NRMSE metric, the Eco-Epi model consistently outperformed the null model throughout all years and outperformed the NB model in all years except 2014, 2019 (0.72 vs 0.20 and 0.66 vs 0.25, ). 
Across all weekly predictions, the Eco-Epi model did not produce outlier results. 
Furthermore, 87 out of 88 weekly predictions ($98.8\%$) fell within the bin corresponding to the highest predicted density (Supplementary 1.2.2.7 and 1.2.2.8).  
Based on AVE values, $100\%$ of all weekly predictions for Orange County had an error of fewer than 4 cases. 
The detailed distribution of the errors is as follows: error $<1$: $36.36\%$; error $<2$: $63.64\%$; error $<3$: $90.91\%$; error $<4$:  $100\%$.  

For Harris County, the Eco-Epi model outperformed both the null model and the NB model in terms of logarithmic score in all years from 2011 to 2024. 
Based on the NRMSE metric, the Eco-Epi model demonstrated superior performance compared to the null baseline in all years and outperformed the NB model in all years except 2019, 2020, and 2023 (0.71 vs 0.38, 0.76 vs 0.07, and 0.59 vs 0.11, respectively) (Supplementary 1.2.2.13 and 1.2.2.14).
In total, the Eco-Epi model did not produce outlier predictions in Harris County.
Of the 99 weekly forecasts generated by the Eco-Epi model from 2011 to 2024, the prediction for 92 weeks ($93\%$) fell within the bins corresponding to the highest predictive density (Supplementary 1.2.2.9 and 1.2.2.10).
Based on AVE values, $94.95\%$ of all weekly predictions for Harris County had an error of fewer than 5 cases, while only $5.05\%$ exhibited an error greater than or equal to 4. 
The detailed distribution of the errors is as follows: error $<1$: $37.37\%$; error $<2$: $68.69\%$; error $<3$: $87.88\%$; error $<4$: $91.92\%$; error $<5$: $94.95\%$; and error $\geq 5$: $5.05\%$. 

\begin{figure}
    \centering
    \includegraphics[width=0.7\linewidth]{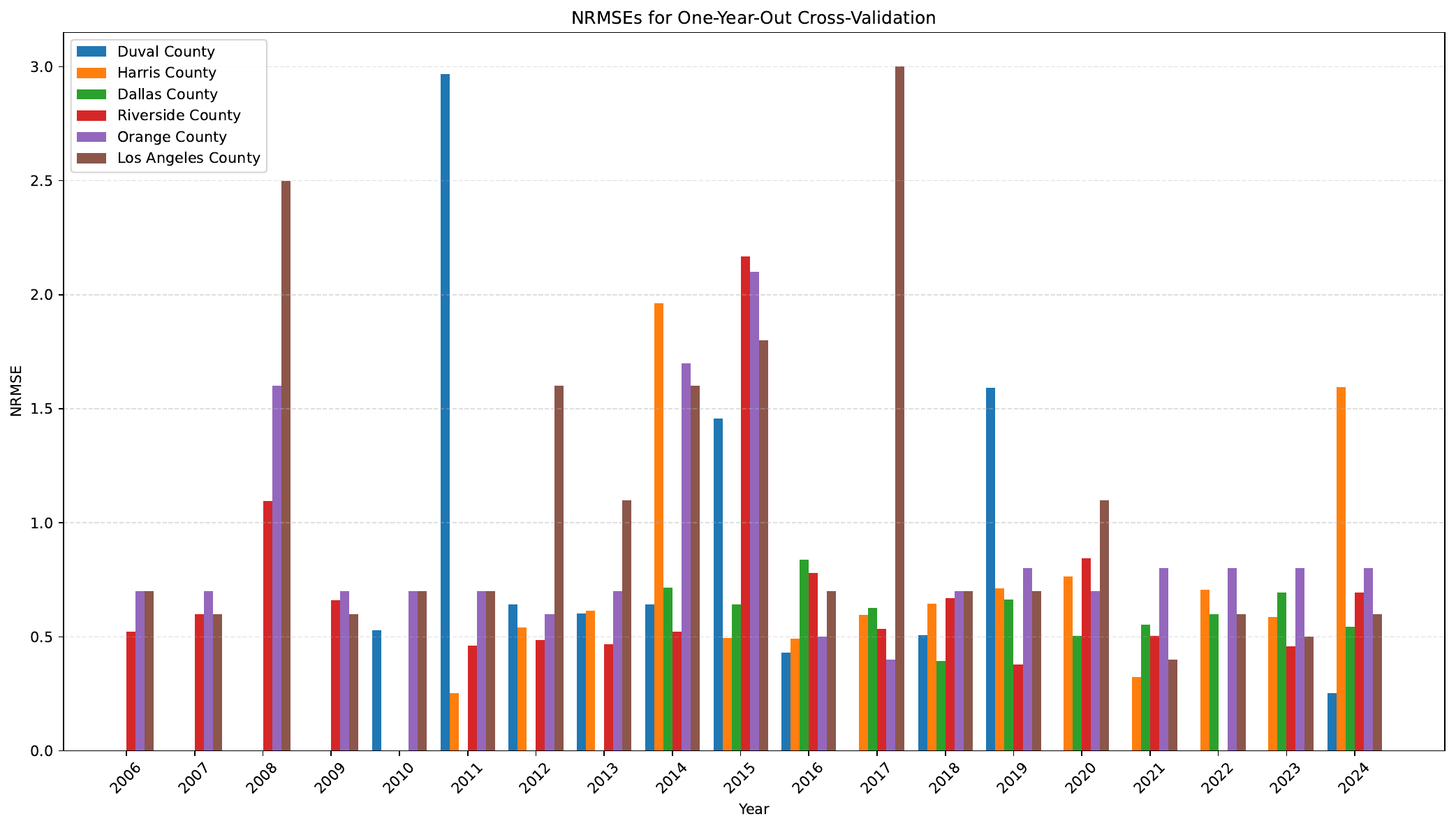}
    \caption{Leave-One-Year-Out NRMSE values in prediction of the severity of the outbreak for Riverside, Orange, Los Angeles, Dallas, Harris and Duval Counties.
    Each value corresponds to the NRMSE for the prediction for the excluded year.}
    \label{fig:RMSE_comparison}
\end{figure}

Figure~\ref{fig:RMSE_comparison} shows the NRMSE for Riverside, Orange, and Los Angeles counties, Dallas, Harris, and Duval counties. 
During the study period, the model yielded NRMSE~$<1$ in 84\,\% of years for Orange County (16 years of 19 years) and in 88\% of years for Riverside County (15 years of 17 years). 
For Los Angeles County $63\,\%$ of the years between 2006 and 2024 (12 of 19) produced NRMSE values below 1.
In the other three counties, the proportion with NRMSE $< 1$ was highest in Dallas County ($100\%$; 11 of 11), followed by Harris County ($86\%$; 12 of 14) and Duval County ($70\%$; 7 of 10).
Regarding Orange County and Riverside County, the lowest forecast accuracy based on NRMSE occurred in 2015 for Orange County and Riverside County when NRMSE reached 2.10 in Orange and 2.16 respectively.
In Los Angeles, the highest NRMSE occurred in 2017 (NRMSE=3.0).
In Dallas County, the least accurate prediction by NRMSE occurred in 2016 (NRMSE=0.84).
In Harris County, the least accurate prediction by NRMSE occurred in 2014 (NRMSE=1.96).
In Duval County, the least accurate prediction by NRMSE occurred in 2011 (NRMSE=2.97).

\subsubsection{Retrospective forecast for risk severity}

For Riverside County, we used data from 2006--2021 and 2006--2023 to generate predictions for 2023 and 2024.
According to the weekly reported human cases, there were 10 weeks with reported cases in 2023.
In terms of the annual sum of the logarithmic score, the Eco-Epi model outperformed the null baseline and the NB model (-1.87 vs -9.69 and -7.48).
Furthermore, the reported case values for each week fell within the bins of the highest predicted probability.
In terms of NRMSE for 2023, the Eco-Epi model outperformed both the null baseline and negative binomial model ( 0.45 vs 3.04 and 0.71). 
For the retrospective forecast of 2024, the data indicate 4 weeks with reported cases (weeks 31, 34, 38, and 40).
In all 4 weeks, the Eco-Epi model substantially outperformed the null model and the Nb model in terms of logarithmic score (-0.73 vs -3.93 and -3.24).
Eco-Epi model had the same behavior in terms of NRMSE and outperformed both models (0.69 vs 1.91 and 0.83).
Furthermore, the reported case counts for each of these weeks fell within the bins associated with the highest predicted probability density, resulting in the highest logarithmic scores (Table \ref{tab:retro-nrmse-logscore} and Supplementary 1.2.3.1-8)).

Regarding Orange County and the retrospective forecast for 2023, 6 weeks of reported human cases were identified. 
The Eco-Epi model outperformed both the null baseline and the negative binomial model on the basis of the annual sum of the logarithmic score (-2.77 vs -4.24 and -3.81). 
All reported case counts fell within the high-density region of the predicted distribution of the Eco-Epi model.  
In terms of NRMSE the performance of the Eco-Epi model was close to null baseline and negative binomial model (0.8 vs 0.5 and 0.7).
The logarithmic score of the Eco-Epi model in this year showed better performance compared to the null baseline and negative binomial model (-0.97 vs -1.48 and -1.35).
NRMSE showed similar performance with null baseline and negative binomial model (0.8 vs 0.5 and 0.5).
Moreover, the reported case count fell within the high-probability region of the Eco-Epi Poisson model (Table \ref{tab:retro-nrmse-logscore} and Supplementary 1.2.3.9-16).

For Los Angeles County (2023), the model outperformed the null model and the NB model when assessed by the annual sum of logarithmic scores (-7.17 vs -11.75 and -19.45).  
Among the weekly forecasts issued for 2023, the prediction for 11 weeks (91.66\%) fell within the highest density region of the predictive Eco-Epi Poisson distribution. 
In terms of NRMSE Eco-Epi model outperformed the null baseline and negative binomial model (0.5 vs 1.1 and 0.7).
For 2024, the model outperformed the null model and the NB model in terms of annual sum of logarithmic score (-5.58 vs. -8 and -10.85).
All forecasts fell within the highest-density region of the Eco-Epi Poisson distribution. 
NRMSE also showed the same outperforming compared with null baseline and negative binomial model (0.6 vs 0.9 and 0.7) (Table \ref{tab:retro-nrmse-logscore} and Supplementary 1.2.3.17-24)

For Duval County, Florida, data from 2010–2018 were used to generate a retrospective forecast for 2019, and data from 2010–2019 were used to generate a retrospective forecast for 2024; No cases were reported in 2017 or 2020–2023.
In both 2019 and 2024, the Eco--Epi model outperformed the null baseline and the NB on both the logarithmic score and the NRMSE.
All prediction values fell in the Eco-Epi Poisson distribution region with the highest density, and there were no outlier predictions (Table \ref{tab:retro-nrmse-logscore} and Supplementary 1.2.3.41-48).

For Dallas County, Texas, data from 2014-2022 and 2014-2023 were used to generate a retrospective forecast for 2023 and 2024 respectively.
For both 2023 and 2024 the Eco-Epi model outperformed both null baseline and NB model in terms of RMSE and logarithmic score.
Across the 18 weekly forecasts for Dallas County (9 in 2023 and 9 in 2024), all reported cases fell within the highest density bins of the Eco-Epi Poisson pmf; consequently, no outlier ($-10$) scores occurred (Tables \ref{tab:retro-nrmse-logscore} and Supplementary 1.2.3.25-32).

For Harris County, Texas, data from 2011-2022 and 2011-2023 were used to generate a retrospective forecast for 2023 and 2024 respectively.
For both 2023 and 2024 the Eco-Epi model outperformed both the null baseline and the NB model in terms of NRMSE and logarithmic score.
Totally there were 12 weeks of prediction without outlier prediction.
Additionally, 10 weeks prediction fell with the highest density bin of the Eco-Epi Poisson distribution (Tables \ref{tab:retro-nrmse-logscore} and Supplementary 1.2.3.33-40).

\begin{table}[htbp]
\centering
\caption{Retrospective forecast performance by county and year: NRMSE and logarithmic score (Eco--Epi, Null, NB).}
\label{tab:retro-nrmse-logscore}
\begin{tabular}{ll
                S[table-format=2.2] S[table-format=2.2] S[table-format=2.2]
                S[table-format=-2.2] S[table-format=-2.2] S[table-format=-2.2]}
\toprule
\multicolumn{2}{c}{} & \multicolumn{3}{c}{NRMSE} & \multicolumn{3}{c}{Log score} \\
\cmidrule(lr){3-5}\cmidrule(lr){6-8}
County & Year & {Eco--Epi} & {Null} & {NB} & {Eco--Epi} & {Null} & {NB} \\
\midrule
Dallas       & 2023 & 0.70 & 7.38 & 2.06 & -2.61 & -13.71 &  -9.13 \\
Dallas       & 2024 & 0.55 & 3.70 & 2.39 & -2.01 & -13.63 & -16.26 \\
Harris       & 2023 & 0.49 & 2.16 & 0.20 & -0.17 &  -1.30 &  -0.73 \\
Harris       & 2024 & 1.59 & 9.55 & 3.60 & -15.58 & -32.33 & -24.80 \\
Duval        & 2019 & 1.65 & 7.12 & 5.43 & -3.10 &  -5.48 & -15.29 \\
Duval        & 2024 & 0.25 & 5.08 & 1.39 & -1.43 &  -7.54 &  -5.01 \\
Orange       & 2023 & 0.80 & 0.50 & 0.70 & -2.77 &  -4.24 &  -3.81 \\
Orange       & 2024 & 0.80 & 0.50 & 0.50 & -0.97 &  -1.48 &  -1.35 \\
Riverside    & 2021 & 0.45 & 3.04 & 0.71 & -1.87 &  -9.69 &  -7.46 \\
Riverside    & 2024 & 0.69 & 1.91 & 0.83 & -0.73 &  -3.93 &  -3.24 \\
Los Angeles  & 2023 & 0.50 & 1.10 & 0.70 & -7.17 & -11.75 & -19.45 \\
Los Angeles  & 2024 & 0.60 & 0.90 & 0.70 & -5.58 &  -8.00 & -10.85 \\
\bottomrule
\end{tabular}
\end{table}

\noindent For all counties, the temporal risk maps for retrospective forecast have been plotted in Supplementary 1.2.3).

\subsection{Long-term climatic trends and spillover risk}
To evaluate whether there are long-term trends in spillover risk, we constructed time series of the number of risky days for each year from 1990 to 2024 and performed statistical trend analyses separately for each county (Section \ref{Climate}).

For Riverside County, linear regression revealed a statistically significant positive trend, with a slope of 1.17 ($p = 0.003$). 
This result was corroborated by Spearman’s rank correlation, which showed a moderate positive association between time and value ($\rho = 0.53$, $p = 0.001$). 
To confirm a nonparametrical monotonic trend, we applied the Mann--Kendall test, which also indicated a statistically significant increase ($p = 0.001$) with a slope estimate of 1.12. 
Together, these findings provide strong and consistent evidence of a long-term upward trend in Riverside County.

Similarly, for Orange County, linear regression indicated a statistically significant positive trend, with a slope of 0.58 ($p = 0.051$). 
The Spearman rank correlation also revealed a moderate monotonic association between time and values ($\rho = 0.38$, $p = 0.027$), corroborating the presence of a consistent upward pattern. 
This finding was further supported by the Mann--Kendall test, which detected a statistically significant increasing trend ($p = 0.026$) with a slope estimate of 0.65. 
Together, these results provide strong evidence for a persistent long-term increase in Orange County during the study period.

For Los Angeles County, the linear regression indicated a statistically significant upward trajectory, with a slope of 0.68 and associated $p = 0.009$. 
This parametric result was corroborated by the non-parametric Spearman rank correlation, which showed a moderate positive association between time and the response variable ($\rho = 0.45,\; p = 0.007$). 
To confirm the presence of a monotonic trend without relying on distributional assumptions, a Mann--Kendall test was applied and also detected a significant increasing trend ($p = 0.014$) with a Sen slope estimate of 0.63.  
Together, these three independent tests provide consistent evidence of a long-term increase in Los Angeles County (Supplementary 1.3).

Consistent with our finding of increased spillover risk in Los Angeles County (2006–2024), Bayles et al. (2004–2021) reported positive monotonic trends in parts of Los Angeles County based on mosquito infection metrics; however, their analysis does not provide county-level trend claims for Orange or Riverside, limiting direct comparison to those counties \cite{bayles2024long}.

\section{Discussion}
This study presents a two-step probabilistic framework to forecast when the WNV spillover begins and how severe it will be, using temperature-driven mosquito dynamics and cumulative temperature. 
Across diverse counties and years, the parsimonious model demonstrates robust out-of-sample skill and yields interpretable daily risk signals that can inform early season planning.
We compare the models performance with baseline models and conclude with limitations and operational implications.

\subsection{Spillover onset}
In general, the onset model delivered consistent and calibrated intervals between sites and years.
Retrospective and cross-validation results suggest that the approach is robust to missing years and climatic heterogeneity. 

Our onset forecasts show the expected trade-off between coverage and precision: as the confidence level \(\alpha\) increases, prediction intervals widen but capture a larger fraction of true onset dates.
This pattern is exactly what a calibrated probabilistic system should exhibit and indicates that the model’s uncertainty quantification is sensible rather than overconfident.
In practice, intermediate confidence levels (\(\approx 0.7\text{--}0.8\)) offer the best operational balance intervals, while success rates are high.
We verified calibration by comparing the empirical coverage \(\beta(\alpha)\) with the nominal \(\alpha\); the coverage increased with \(\alpha\) and closely followed it.

National and state surveillance shows that WNV activity typically spans summer to fall, with human cases peaking in late August to early September \cite{cdc2022westnile,CDC_WNV_Guidelines_2024}.
In southern California, first detections and cases likewise cluster in mid–late summer; for example, Los Angeles County often reports initial onsets from late July to late August \cite{lacdph_home,CDPH2024WNV}. Consistent with this seasonality, our predicted onset windows for Orange, Riverside, and Los Angeles fall within these empirically expected periods (Supplementary Sec.~1.1.1), indicating alignment across contrasting coastal–inland climates.

The detected pattern in predictive PDF's for spillover, which was explained in Section~\ref{Predictive PDFs}
probably reflects greater climatic heterogeneity and ecological diversity in Los Angeles and Riverside than in Orange.
By climatic heterogeneity, we mean within-county variation in pre-onset cumulative temperature $T$ (e.g., range/IQR across onset days); by ecological diversity, landscape, and hydrological characteristics that shape mosquito habitats.

Considering the Köppen climate classification, Orange County (warm Mediterranean, Csa/Csb) is strongly moderated by the marine layer and is therefore more thermally uniform; the pre-onset \(T\) lies in a narrow band and the onset windows are shorter.
Los Angeles spans coastal Mediterranean (Csb/Csa) to inland semi-arid/desert (BSh/BWh) climates, with elevation contrasts and urban heat-island effects that broaden the distribution of pre-onset \(T\) and produce wider onset windows.
Riverside similarly transitions from Mediterranean foothills (Csb/Csa) to extensive semi-arid/desert (BSh/BWh) along the Perris–San Jacinto corridor and the Coachella Valley; irrigated agriculture and riparian corridors further diversify habitats without narrowing the thermal spread.
This interpretation is consistent with the Köppen–Geiger maps for southern California (coastal Csa/Csb vs. inland BSh/BWh) \cite{kesseli1942climates,beck2006characterizing,beck2018present} and with the predictive PDF for spillover onset, where the pre-onset range of \(T\) is tightest in Orange and wider in Los Angeles and Riverside.
Operationally, running the model on subcounty scales in Los Angeles and Riverside should tighten the onset intervals while preserving calibration.
This interpretation is consistent with Bayles et al. (2004-2021), who found persistent and oscillating WNV hot spots in the Los Angeles metropolitan area and attributed this instability to climatic and ecological variability (e.g., topographic variation, temperature fluctuations, rainfall anomalies, ENSO) \cite{bayles2024long}

In our spillover onset model, onset timing relies solely on the predicted abundance of mosquitoes $M$ and the cumulative temperature $T$; unmodeled drivers (e.g., precipitation, host / community shifts) may widen the intervals in complex settings.
We also approximate future temperature with 5-year means, pragmatic and consistent with prior practice, but this smooths interannual extremes that can determine exact onset timing.
Together, these points suggest a clear operational pathway: deploy the model with intermediate confidence; interpret interval width through local thermal variability; and, where feasible, downscale to subcounty units to reduce heterogeneity. 
Quantitative details are given in the tables and Supplementary Figures. Bottom line: the onset component delivers calibrated, actionable windows whose width is ecologically interpretable and tunable to decision needs.

\subsection{Spillover severity}
After determining the onset, we quantified the severity with the rate surface of the Eco-Epi Poisson model $\lambda(M,T)$, which links the abundance of temperature-driven mosquitoes $M$ and cumulative thermal exposure $T$ to the expected daily human cases.
This complements the PDF of the onset by specifying the intensity of daily spillover after the onset using the joint drivers $(M,T)$. We then forecast expected daily cases using this surface (Section~\ref{ratefunction}).

\subsubsection{Rate surface of spillover}
Across all counties, Eco-Epi rate surfaces share several common characteristics.
In Section~\ref{sec: rate} we showed that the estimated Eco-Epi spillover rate surface
\(\lambda(M,T)\) crosses the unity isocline (\(\lambda=1\)) as mosquito abundance \(M\)
increases from its minimum to approximately the first quartile of its observed range.
Crossing this unity contour is a statistical action threshold: Inside this unity contour, the implied probability of at least one reported human case is
$\Pr\{N(M,T) \ge 1\} \ge 1-e^{-1}=0.63$ (see the red dashed contour in Figure \ref{fig:intermediate_M_OR}).

This threshold is different from the mechanistic epidemic threshold \(R_0>1\) or \(R_t>1\);
it is related to the expected case rate conditional on \((M,T)\) rather than to a reproduction
number.
Operationally, the unity contour functions as an action threshold analogous to entomological risk thresholds used in practice, such as the Vector Index (infected mosquitoes per trap night) and the California Mosquito-Borne Virus Risk Assessment (CMVRA) \cite{VectorSurvRiskAssessment2025}. 
Consistent with a practical threshold in vector density, human spillover is rarely observed when mosquito abundance remains below a lower range, underscoring the importance of achieving critical vector densities for human WNV outbreaks. 
This pattern is supported by the 2010 Maricopa County, Arizona, investigation, in which gravid traps captured significantly more \textit{Cx. quinquefasciatus} in outbreak neighborhoods than in a demographically matched area without human cases \cite{godsey2012entomologic}.
Likewise, in northeastern Colorado (2006-2007), time series analyses showed that weeks with human cases were preceded by elevated \textit{Culex} abundance (on the order of several weeks), consistent with a practical abundance threshold \cite{bolling2009seasonal}.
One key point is to keep the abundance of mosquitoes below this critical threshold through different mitigation strategies, acting on the size of the mosquito population. 

In addition, the Eco-Epi spillover rate surface shows that the highest spillover values cluster in the intermediate window of $T$, with markedly lower rates at both high and low values of $T$.
Because these intermediate values of normalized cumulative temperature $T$ occur around $20$ --$30\,^\circ\mathrm{C}$, the pattern is consistent with evidence that WNV transmission peaks near $23$ --$26\,^\circ\mathrm{C}$ and declines at cooler or hotter temperatures \cite{shocket2020transmission,mordecai2019thermal}.
Mechanically, 
cool temperatures reduce mosquito bite and viral replication and
lengthen the extrinsic incubation period (EIP), suppressing transmission and favoring reduced activity
\cite{reisen2006effects, dohm2002effect}; at high temperatures, adult survival declines and, for some
vector–virus combinations, vector competence is reduced, again lowering the transmission
potential \cite{shocket2020transmission, mordecai2019thermal,ciota2014effect}.
Finally, within this suitable temperature window, the Eco--Epi spillover rate increases with mosquito abundance up to a threshold and then declines, producing a hump-shaped response. 
This pattern is consistent with density-dependent constraints (carrying capacity) in mosquitoes \cite{agnew2000effects,ower2019effects,marini2016role} and with evidence that the transmission of WNV peaks at intermediate temperatures and decreases under cooler or hotter conditions \cite{shocket2020transmission,mordecai2019thermal}.

As illustrated by the Eco-Epi rate surfaces and seasonal trajectories $(M,T)$ in the six counties, late–winter and spring mosquito abundance is a key determinant of outbreak severity. 
When the early season $M$ is large, the trajectory enters the elevated–$\lambda$ region sooner and remains there longer, increasing the cases.
For Orange County, we illustrate this effect by increasing or decreasing winter–spring temperatures; the resulting changes in the epidemic trajectory are shown in Supplementary~4.
In practical terms, a warm winter and a moderate (mild/wet) spring can boost $M$ before summer, positioning the system to begin the transmission season at greater risk.
This pattern is confirmed by some other field studies \cite{ward2023spatially, brown2015projection}.

\subsubsection{Leave-one-year-out cross-validation and retrospective evaluation}
In both leave-one-year-out cross-validation and retrospective forecasts, the Eco-Epi model achieved low NRMSE and favorable log scores across sites (Supplementary~1.2.2.13--1.2.2.14), including high-burden seasons: Duval 2011-2012; Orange 2008, 2014-2015; Dallas 2016-2021; Harris 2014, 2018, 2024; Los Angeles 2008, 2011-2017, 2020; Riverside 2008, 2013, 2015, 2017.
The performance of the Eco-Epi model exceeded the null baseline and the negative binomial model without degradation in the years of high incidence, and the predictive distributions were well calibrated (Fig.~\ref{fig:RMSE_comparison}, Table~\ref{tab:retro-nrmse-logscore}, Supplementary 1.2.2 and 1.2.3).
The model’s day-level forecasts provide the temporal granularity needed for mid-season decision support.

\subsubsection{Per-County Risk Forecasts, Mechanisms, and Mitigation}
For each county, we (i) summarize the prediction for peak time and seasonal intensity; (ii) interpret the surface of the rate $\lambda(M,T)$ to identify mosquito-temperature drivers of risk; (iii) benchmark performance against null baseline and the NB model; and (iv) translate signals into timed, targeted surveillance, and control. 
These points explain what the model predicts, how well it performs, and how agencies can use it throughout the season.
While in the following we provide an overview, detailed, county-specific explanations are provided in the Appendix J.

Duval County shows year-round transmission potential: Our Eco-Epi model detects early‐season risk that intensifies into summer and fall, with a primary window from late July to early November.
Based on the rate surface of the Eco-Epi model, the spillover peaks when the abundance of moderate to high mosquitoes coincides with intermediate normalized cumulative temperatures,\textbf{ $T \in[10,18^\circ$C]}.
The risk was predicted even in years without reported human cases, consistent with independent activity indicators and possible pandemic era under-measurement.
The model outperforms the null baseline and negative binomial model and provides actionable, day‐level guidance for surveillance and control (Supplementary 1.2.2.15).

In Dallas County, the Eco-Epi model recovers a robust bimodal spillover pattern with a primary mid-summer peak (mid-July to mid-August) and a secondary early fall rise (mid-September to mid-October), consistent with surveillance \cite{DCHHS_Arbovirus_Surveillance_Dallas}.
The highest predicted risk occurs in the windows of $T\in [8,18~^\circ C$].
Thermal forcing likely drives bimodality: Extreme summer heat suppresses activity, with a resurgence as conditions cool, supporting two timed control surges: late June to early July (approximately 2 to 3 weeks ahead of the usual mid-July to mid-August peak) and again in late September to mid-October when late summer heat breaks following fall rainfall (Supplementary 1.2.2.15).

For Harris County, the rate surface of the Eco-Epi model shows that the highest spillover occurs within the window of \textbf{$T \in [4,19.45~^\circ C]$ }.
The model recurrently predicted the risk peak from late summer to early fall.
In the severe outbreak of 2014, the forecasts aligned with the surveillance: mosquito pools first tested positive in early June, peaked in late July (week~30); human cases peaked in mid-August, with elevated activity through mid-October.
Operationally, routine surveillance and control should begin in late spring and continue through late fall, in accordance with CDC season-long guidelines \cite{CDC_WNV_Surveillance_Control_2024} (Supplementary 1.2.2.15).

In Orange County, the Eco-Epi model shows strong skill with a stable seasonal pattern: predicted risk increases in late July, concentrates in late summer to early fall, and decays through October, consistent with county surveillance \cite{OCHCA_WNV_2011_2018,OCMVCD_EmergencyPlan_2015}.
The Eco-Epi rate surface peaks at moderate cumulative temperatures ($T\in [10,16^\circ C$]). 
After onset, severity is governed by how quickly $M$ exceeds a threshold, persistence of $T \in[11,14\,^\circ C]$, and dwell time in the high-$\lambda$ contour (e.g., $\ge$3--4 consecutive weeks). 
The prediction of risk for heavy seasons shows broad high-intensity blocks, including unusually late activity that the model captures (Supplementary 1.2.2.4); moderate seasons exhibit shorter and patchier predicted risk (Supplementary 1.2.2.15).

Regarding Riverside County, the Eco-Epi model reveals a stable high contour $\lambda$ with higher mosquito abundance with normalized cumulative temperatures window $T\in [8,17 ^\circ C]$; large outbreaks occur when seasonal $(M,T)$ trajectories dwell within this high-rate contour through late summer--early autumn. Phenology is recurrent: onset in late August, peak in September, and decline in late October, is consistent with field studies \cite{snyder2020west}.
In heavy season forecasts, the $(M,T)$ trajectory tracks the core of the high-$\lambda$ region, whereas lower activity years intersect it only briefly or late.
The rate surface of the Eco-Epi model shows that the severity after onset is governed by the speed of the rise $M$, the persistence of warm conditions into the early fall and the dwell time in the high $\lambda$ band (Supplementary 1.2.2.15).

For Los Angeles County, the rate surface of the Eco-Epi model indicates that spillover peaks at higher mosquito abundance than in neighboring counties, yet non-trivial risk appears even at moderate levels, consistent with late-season spillovers.
In addition, a recurrent high-risk high risk contour occurs at high $M$ with warm seasonal temperatures ($T\in(12-16)^\circ C$); the trajectory of ($M,T$) in the heaviest seasons enters this high-risk contour early and dwells longer, producing broad September-October windows with occasional November tails.
For Los Angeles County, the average accuracy is modest; Eco-Epi model predictions improve on null baseline and negative binomial model on average but not consistently - probably because the county scale $(M,T)$ blurs strong coastal–interior heterogeneity (elevation and urban heat island effects).
Operationally, subcounty fits should sharpen timing and calibration (Supplementary 1.2.2.15).

Overall, the two-step framework delivers calibrated onset windows and accurate severity forecasts across diverse counties, outperforming baseline models in retrospective and cross-validation tests. The resulting daily risk signals and \(\lambda(M,T)\)–based thresholds are interpretable and operational, supporting dependable early-season planning and season-long surveillance and control. Performance may soften at coarse scales in climatically heterogeneous settings (e.g., Los Angeles), but subcounty deployment can sharpen timing while preserving calibration.

\bibliography{sample}

\section*{Author contributions statement}
\textbf{Hosseini, Saman}: Conceptualization, data curation, formal analysis, investigation, methodology, validation, visualization,
writing – original draft, writing – review \& editing.\\
\textbf{Cohnstaedt, Lee}: Conceptualization, methodology, resources, supervision, validation, writing – original draft, writing, review
\& editing.\\
\textbf{Marjani, Matin}: Data curation, validation, visualization, writing – review \& editing.\\
\textbf{Scoglio, Caterina}: Conceptualization, formal analysis, funding acquisition, methodology, project administration, supervision,
validation, writing original draft, writing, review \& editing.
\section*{Funding}
This work was funded by the USDA National Institute
of Food and Agriculture award number 2022-67015-38059 via
the NSF/NIH/USDA/BBSRC/BSF/NSFC Ecology and Evolution of Infectious Diseases program, and the United States
Department of Agriculture ARS under agreement number 58-3022-1-010.

\end{document}